\documentclass{aa}
\usepackage{graphicx}
\usepackage{txfonts}
 \newcommand{\gesssim}{\mathrel{\hbox{\rlap{\hbox{\lower4pt\hbox{$\sim$}}}\hbox{$>$}}}}

 \newcommand{\teff}{$T_{\rm eff}$}
 \newcommand{\nli}{$\log n$(Li)}

\begin{document}
   \title{The effect of heavy element opacity on pre-main sequence Li depletion}

%   \subtitle{}

   \author{P. Sestito\inst{1,2} \and S. Degl'Innocenti\inst{3,4} \and P.G. Prada Moroni\inst{3,4} \and S. Randich\inst{2}}
 
    \offprints{P. Sestito, email:sestito@arcetri.astro.it}
 
 \institute{INAF/Osservatorio Astronomico di Palermo ``Giuseppe S. Vaiana'',
 Piazza del Parlamento 1, I-90134 Palermo, Italy
 \and 
 INAF/Osservatorio Astrofisico di Arcetri, L.go E. Fermi 5,
             I-50125 Firenze, Italy
 \and
 Dipartimento di Fisica ``E. Fermi'', L.go B. Pontecorvo 2,
              I-56127 Pisa, Italy
 \and INFN, Sezione di Pisa, L.go B. Pontecorvo 2,
              I-56127 Pisa, Italy}

 \titlerunning{Heavy element opacity and Li depletion}
 \date{Received Date: Accepted Date}

% \abstract{}{}{}{}{} 
% 5 {} token are mandatory
 
  \abstract
  % context heading (optional)
  % {} leave it empty if necessary  
   {Recent 3-D analysis of the solar spectrum data 
suggests a significant change of
the solar chemical composition. This may affect the temporal evolution of the
surface abundance of light elements since the extension of the convective
envelope is largely affected by the internal opacity value.
}
  % aims heading (mandatory)
   {We analyse the influence of the
adopted solar mixture on the opacity in the convective envelope of
pre-main sequence (PMS) stars and thus on PMS  lithium depletion. The surface
Li abundance depends 
on the relative efficiency of several processes,
some of them still not known with the required precision; this paper 
thus analyses one of the aspects of this ``puzzle''.}
  % methods heading (mandatory)
   {Focusing on PMS evolution, where the largest amount of Li
burning occurs, we computed stellar models for three selected masses (0.8, 1.0
and 1.2 $M_{\sun}$, with $Z=0.013$, $Y$=0.27, $\alpha$=1.9) by varying the
chemical mixture, that is the internal element distribution
in $Z$. We analysed  the
contribution of the single elements to the opacity 
at the temperatures and
densities of interest for Li depletion. 
Several mixtures were obtained by varying the abundance of the most important
elements one at a time; we then calculated the corresponding PMS Li
abundance evolution.
   }
  % results heading (mandatory)
{We found that a mixture variation does change the Li abundance:
at fixed total metallicity, the
Li depletion increases when increasing the fraction of elements heavier
than O.} 
  % conclusions heading (optional), leave it empty if necessary 
   {}

\keywords{Stars: abundances
-- Li -- Stars: Evolution}

\maketitle

%
%________________________________________________________________

\section{Introduction}\label{intro}
 In the last twenty years several observational studies of 
lithium ($^{7}$Li) in stars have been performed, many
 of which focus on open cluster objects in the 
zero-age main sequence (ZAMS) and in the
main sequence (MS) phase (see the reviews
by Jeffries \cite{jef00}; Randich \cite{R05_cast04}; and 
the references quoted in Table~1 of Sestito \& Randich
\cite{SR05}).
The temporal evolution of the surface Li
 abundance has also been investigated theoretically (e.g. D'Antona \& Mazzitelli \cite{dm97}; review by
Pinsonneault
 ~\cite{pinso} and references therein; Siess, Dufour \& Forestini
\cite{siess}).  The
 interest in Li depletion in stars comes from
the fact that, due to its relatively low burning temperature ($\sim
 2.5\times$10$^{6}$ K), Li is a key tracer of the efficiency of 
 mixing processes in stellar envelopes.

In spite of great efforts, the theory is still unable to reproduce the temporal
behavior of surface Li abundances during the pre-main Sequence (PMS) 
and MS phases, i.e. model predictions do not match
the measured abundances for the Sun and for stars in open clusters of
different ages (see, e.g.  
Jeffries \cite{jef00} and references therein).

The task is not easy because the temporal evolution of the surface chemical
abundance (when the accretion phase has ended) depends on the relative
efficiency of several processes: mixing, microscopic diffusion, radiative
levitation and mass loss.  In particular, the abundances of the light elements,
which are easily burned, are extremely sensitive to the extent of 
envelope mixing.  There are two main sources of uncertainties in
these processes: the first
one related to macroscopic processes and the other to microscopic physical
ingredients adopted.  Regarding the former, it is well known that due to the
lack of a self-consistent treatment of convection, stellar modelers are
compelled to use relatively crude approximations depending on free parameters, such as
the mixing length formalism.  Even a precise physical treatment of the efficiency of the mass
loss is not available.  The situation is even worse if the assumption of spherical symmetry is
relaxed allowing for rotation, or if magnetic
fields are present.
 Once a scheme has been chosen for the treatment of a given macroscopic
 process (i.e. convective mixing), the prediction of its efficiency is still
 affected by uncertainties in the basic physical inputs 
 such as the nuclear reaction rates, the equation of state and the opacity of the
 stellar matter (see e.g., D'Antona \& Mazzitelli ~\cite{dm97}; Pinsonneault
 ~\cite{pinso}; Turcotte \& Christensen-Dalsgaard \cite{Turcotte};
 Palla~\cite{Palla}; Jeffries \cite{jef_cast04} for some results of
 evolutionary models). 
The canonical scenario assumes spherical symmetry with constant mass
 and adopts the classical Schwarzschild criterion for the identification of
 convective boundaries. 

A fundamental role in Li evolution
is played by uncertainties in the radiative opacity (which affects the
 depth of the convective envelope) and thus by
 the metal content, both the global abundance (i.e. the metallicity) and the
 distribution of the elements in the mixture.
Recent analysis of solar spectroscopic data using 3-dimensional hydrodynamic
 atmospheric models (see Asplund et al.  \cite{asplund2004}, Asplund, Grevesse,
 \& Sauval \cite{asplund2005} -- hereafter As04, As05 -- and references
 therein) have reduced the derived abundances of CNO and other heavy elements
 with respect to previous estimates (Grevesse \& Sauval 1998, hereafter GS98).
 GS98 improved the mixture by Grevesse \& Noels (1993, hereafter
 GN93), widely adopted in the literature, revising the CNO and Ne abundance
 and confirming the very good agreement between the new photospheric and
 meteoric results for iron. The $Z/X$ solar value decreased
 from the GN93 value $(Z/X)_{\odot}$=0.0245 to the GS98 estimate
 $(Z/X)_{\odot}$=0.0230 and then to the values $(Z/X)_{\odot}$=0.0176, 0.0165
 found by As04 and As05, respectively. The corresponding solar 
metallicity values (i.e. assuming [Fe/H]=0, see also Sect.~\ref{fixedFe}) 
for the different mixtures
are $Z\sim$0.018 (GN93 and GS98), $Z\sim$0.013 (As04) and $Z\sim$0.012 (As05).
However, the problem of a precise
 determination of the solar mixture is still affected by theoretical and
 observational uncertainties and the quoted error on the $Z/X$ solar value is
 still about 10\% (see e.g. Bahcall \& Serenelli 2005).  

In this paper we analyse the influence of the adopted solar mixture on the
 opacity at the base of the convective envelope of pre-main sequence stars and thus
 on Li abundance in the convective envelope. 
We focus on PMS since,
according to classical models,
the largest amount of surface Li destruction for low mass stars
occurs during this stage.
The results are understood in terms of the contribution of each element
to the total opacity at the physical conditions typical of the bottom
of the convective zone during PMS.
Clearly this is not an attempt to solve
 the ``Li problem'', which is much more complex, but a quantitative
 investigation of one of the aspects of the ``puzzle''.

 A few studies have already been carried out concerning the
 dependence of Li depletion on element abundances and/or on the surface
 stellar opacities for solar-type stars (see e.g. Swenson et al.~\cite{swenson94a};
 \cite{swenson94b} and references therein; 
Piau \& Turck-Chi\`eze \cite{ptc02}; and Piau,
Randich \& Palla \cite{piau}). Swenson et al. explored the effect 
of the opacity values (and of other physical inputs) on
 PMS Li depletion,
 as well as the effects of a variation of the oxygen abundance, trying
 to match the Li abundances observed for the Sun and the Hyades
(under the assumption of no MS depletion), finding
that an enhancement
in [O/Fe]
 leads to a larger Li destruction. They also noted that heavier species 
such as Fe, Ne, Mg and Si might
 significantly contribute to the opacity.

Similarly, Piau \& Turck-Chi\`eze (\cite{ptc02}) presented PMS models with
Pleiades (solar) and Hyades-like compositions.  They investigated the
dependence of Li depletion on metal and He fractions, by varying the global
mixture (and changing the total $Z$); they showed that Li abundances are very
sensitive to composition, suggesting that the dispersion in the observed
values in open clusters could be related to small changes in the metal
mixture among the cluster stars.

Piau et al.~(\cite{piau}) 
investigated the sensitivity of the extension of the
 convective zone during the MS for a mass of $\sim$ 1 M$_{\odot}$, with
 respect to the surface content of CNO. Their models are calculated by
 changing [CNO/Fe] while maintaining [Fe/H] and the effective temperature
 unaltered. Thus, the total $Z$ increases if CNO increases (and vice versa).
 Under these conditions, Piau et al.~found that an enhancement of CNO results
 into a smaller amount of
 Li depletion.
Piau (\cite{piau05})
investigated the history of Li isotopes 
in Population {\sc ii} dwarfs, finding a correlation between
the scatter in Li abundance and variations in [Fe/O]
ratios, i.e. changes in [Fe/O] affect Li depletion more than
a change in the global metallicity.
 
The paper is organized as follows: Sect.~2 briefly
describes the evolutionary code and the physical inputs adopted for the 
calculations; in Sect.~3 we present the results of our
computations for models with different mixtures and fixed metallicity
($Z=0.013$),
which are interpreted in Sect.~4. In  Sect.~5 we also
 discuss the effect of changing both
the mixture and the total metallicity value $Z$. A summary (Sect.~6) closes the paper.

 \section{Evolutionary code and input physics}\label{code}
  
The models have been computed with an updated version of the FRANEC
evolutionary code (see e.g. Chieffi \& Straniero \cite{chieffi}); the physical
inputs adopted for the calculations have been listed in the Appendix of
Cariulo, Degl'Innocenti \& Castellani (2004) -- see also Ciacio,
Degl'Innocenti \& Ricci (1997).  With respect to Cariulo et
al.~(\cite{cariulo}) we updated the equation of state (OPAL
EOS\_2001\footnote{http://www-phys.llnl.gov/Research/OPAL}, see also Rogers
(2001),
Iglesias \& Rogers (\cite{OPAL}) and the conductive opacity (Potekhin \cite{Potekin99}).
The nuclear cross sections for Li burning are taken from the NACRE compilation
(Angulo et al. 1999). The adopted initial value for Li in disk stars is
\nli=3.3\footnote{Adopting the usual scale: \nli=12+$\log
(N_{\mathrm{Li}}/N_{\mathrm H}$).} (see, e.g. Jeffries \cite{jef_cast04}),
or $N_{\mathrm{Li}}/N_{\mathrm H}\sim{2\times{10^{-9}}}$.

 The opacity values are fundamental for model
computation; we adopt the
 radiative opacities by the Livermore group (Iglesias \& Rogers \cite{OPAL})
 for temperatures higher than 12000 K and those by Alexander \& Ferguson
 (\cite{AF94}) for the lower temperatures.  As it will be discussed in the next
 section, we used several different mixtures for the high temperature
 opacities, while the Grevesse \& Anders (\cite{grevesse91}) 
mixture has been adopted
 for the molecular atmospheric 
opacities. We also notice that the adopted EOS 
is calculated for the Grevesse \& Anders (1991) mixture; to our knowledge
EOS tables with an updated mixture are not available, however, as stated
in the OPAL EOS web site, these abundance changes should have very small
effects on the EOS.
 
To determine the extension of the convective envelope the classical 
 Schwarzschild criterion has been adopted and the efficiency of the 
 energy transport has been modeled with mixing length
 formalism. We did not include rotation.
  
Element diffusion is included (Ciacio et al.~\cite{ciacio97}, 
 with diffusion coefficients from Thoul, Bahcall \& Loeb \cite{thoul}), while
 radiative acceleration has been neglected since we analysed only the PMS
 phase of low mass stars ($M\leq$1.2 $M_{\odot}$).
In fact, as discussed by Turcotte, Richer \& Michaud (\cite{turcotte98a})
and Turcotte et al.~(\cite{turcotte98b}),
for stars with solar composition and masses around 1 $M_{\odot}$
the effect of the radiative levitation becomes comparable to that of
the gravitational settling only at the end of the MS phase.
  
In the following calculations, due to the uncertainties already discussed, it
is not possible to perform precise evaluations of the absolute Li surface
values; we will instead perform a ``differential'' analysis with respect
to a ``standard'' model, investigating the effects of a mixture change.  Our
``standard'' model adopts [Fe/H]=0 with the As04 mixture (corresponding to a
metallicity of $Z\sim$0.013), $Y$=0.27 and the mixing length parameter
($\alpha$) equal to 1.9. The effects of a variation of [Fe/H], $Y$ and $\alpha$
within a plausible range of uncertainty will also be discussed.
 
%============================================== TABELLA
\begin{table*}[!] \footnotesize
\caption{Summary of the models
computed in this work. In all cases the total metallicity
is fixed to $Z=0.013$, with the exception
of model no.~16; the PMS evolution has been followed for
the three masses 1.2, 1.0 and 0.8 M$_{\odot}$.}\label{tabella}
\scriptsize
\begin{tabular}{ccccccc}
\hline
\hline
No. & Label & Mixture & Element, opacity, or metallicity variations \\
\hline
1&As04 (standard) &Asplund et al.~(2004) &Standard solar mixture \\
2&GN93 & Grevesse \& Noels (1993)&Standard solar mixture\\
3&GS98 &  Grevesse \& Sauval (1998)& Standard solar mixture\\
4&As05 &Asplund, Grevesse \& Sauval 2005 &Standard solar mixture \\
5&As04 (atmospheric opacity decr. by 5 \%) &Asplund et al.~(2004) &Atmospheric opacity
decreased by 5 \% \\
6&As04 (atmospheric opacity decr. by 10 \%) &Asplund et al.~(2004) & Atm. opac.
decreased by 10 \% \\
7&Only C &-- & Metallicity composed by 100\% C\\
8&Only O &-- & 100\% O\\
9&Only Ne & --&100\% Ne \\
10&C($-$20\%) O(+) &Asplund et al.~(2004) &C mass fraction decreased by 20\%
(compensated by O enhancement) \\
11&O($-$20\%) C(+) &Asplund et al.~(2004) &O mass fraction decreased by 20\%
(compensated by C enhancement) \\
12&Ne(+30\%) O($-$) &Asplund et al.~(2004) &Ne mass fraction enhanced by 30\%
(compensated by O decrease) \\
13&Ne($-$30\%) O(+) &Asplund et al.~(2004) &Ne mass fraction decreased by 30\%
(compensated by O enhancement) \\
14&Heavy elements (+20\%) &Asplund et al.~(2004) &Elements heavier than O
increased by 20\% \\
15&CNO (+20\%)&Asplund et al.~(2004) &CNO increased by 20\% \\
16&GN93 $Z=0.018$ & Grevesse \& Noels (1993)&$Z=0.018$, standard solar mixture\\
\hline
\hline
\end{tabular}
\end{table*}
 
 %============================================== FIGURA 1
 \begin{figure}
 %\label{Modelli}
 %\resizebox{0.75\textwidth}{!}{
 \includegraphics[width=7cm]{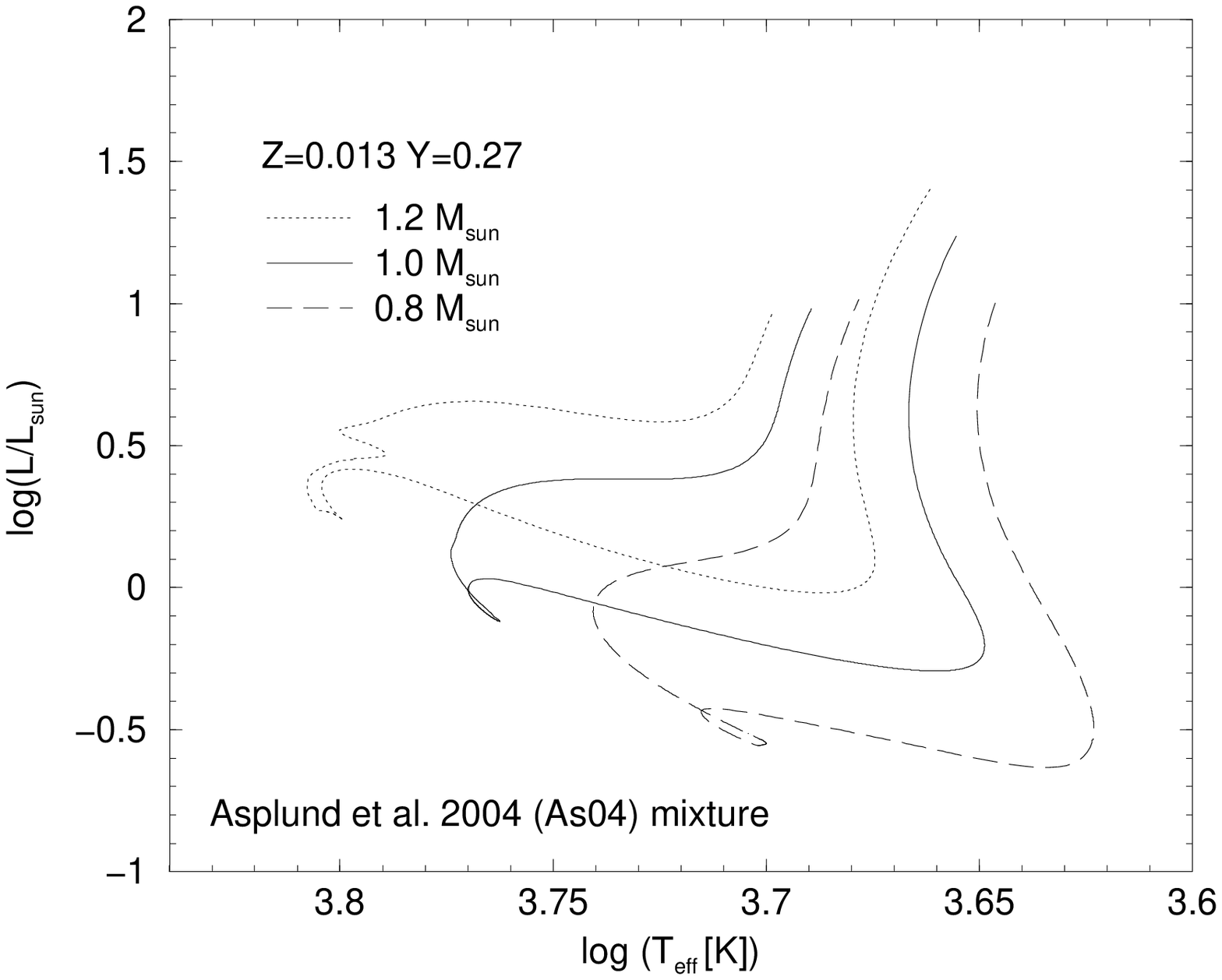}
 
 \includegraphics[width=7cm]{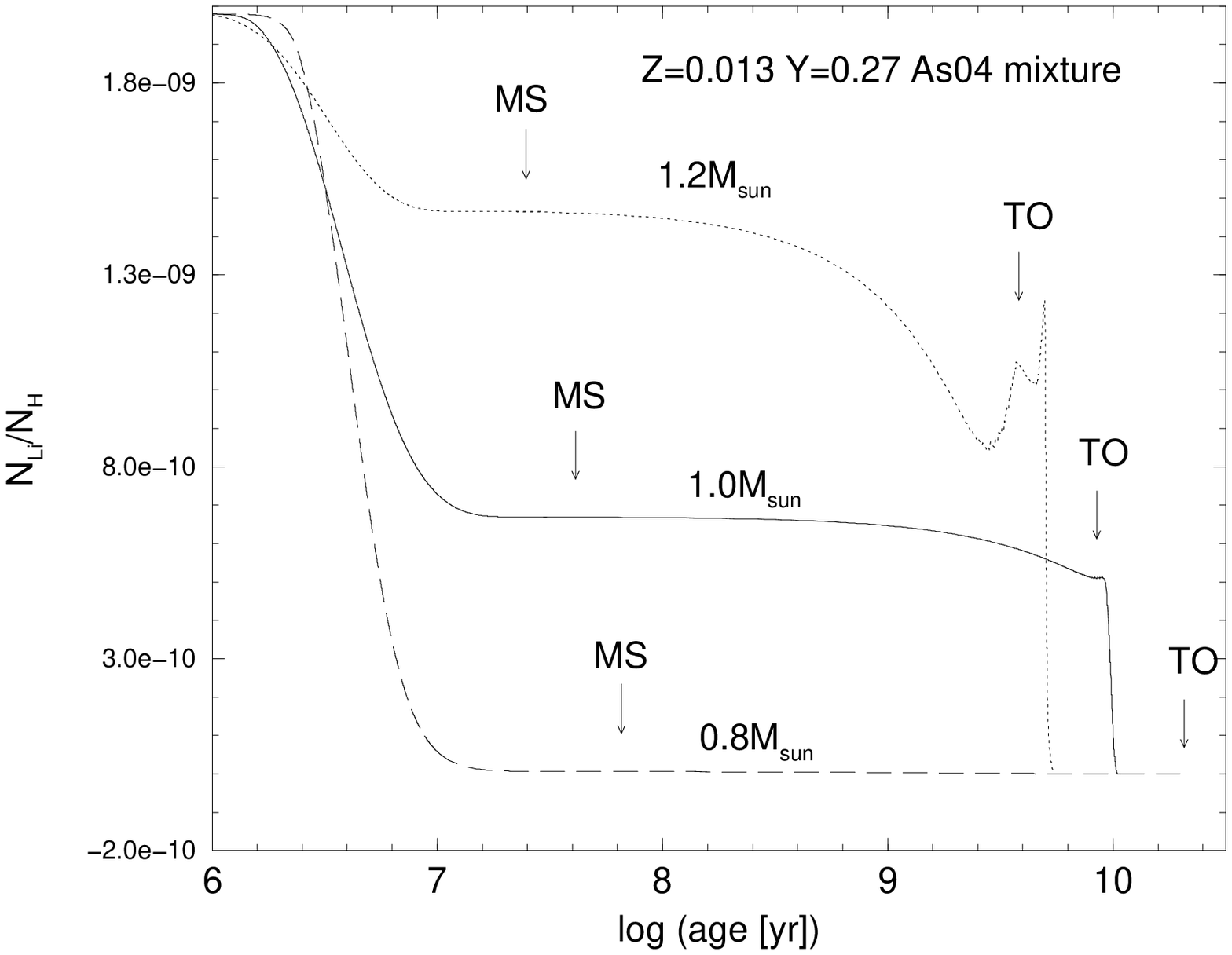}
 
 \includegraphics[width=7cm]{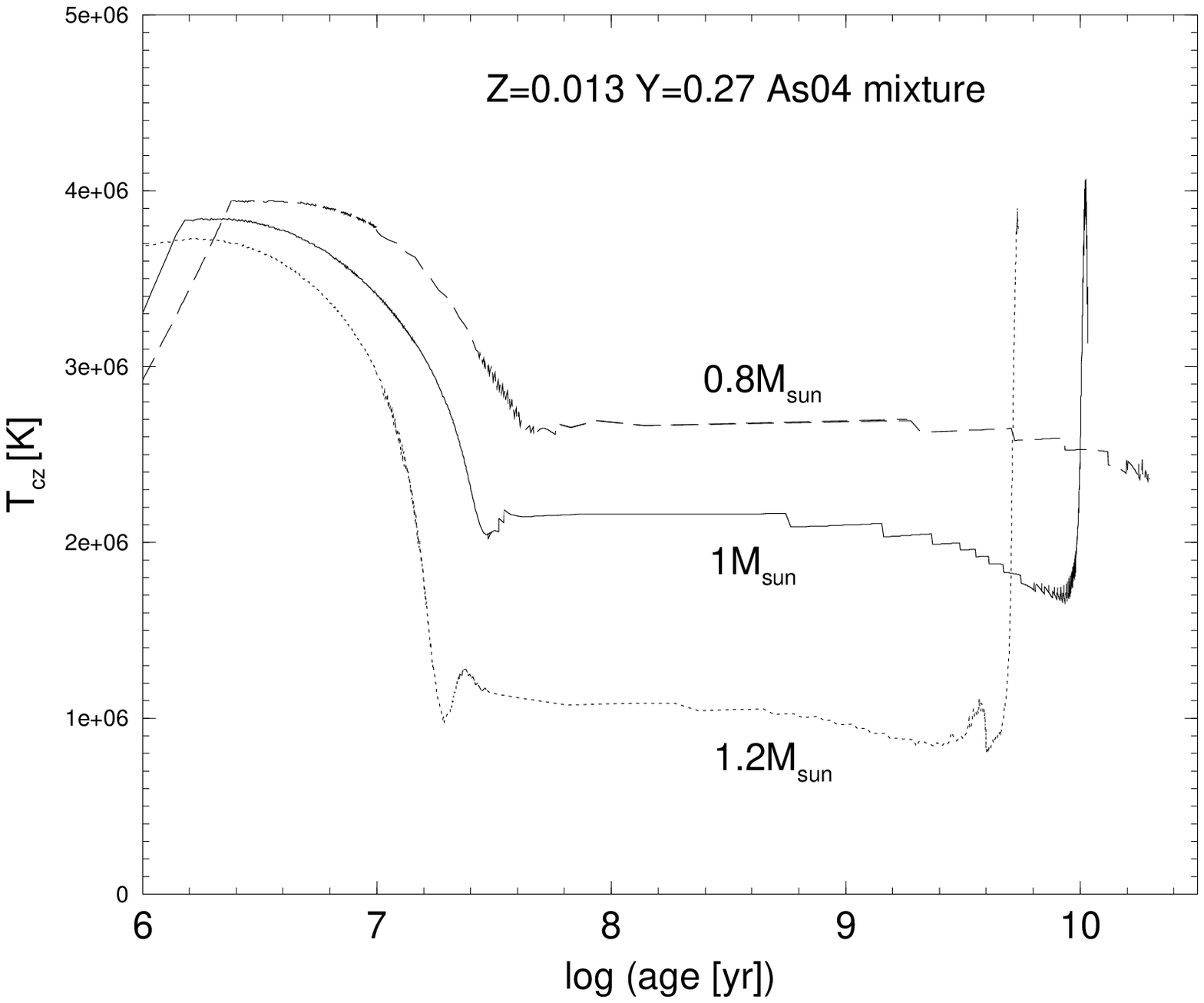}
 \caption{Upper panel: the evolution in the HR diagram of our three selected
   ``standard'' models (0.8, 1.0 and 1.2 $M_{\odot}$, with
$Z$=0.013, $Y$=0.27 and
   $\alpha$=1.9; Asplund et al.\,2004 -- As04 -- mixture) from the PMS to the RGB phase.
 Middle panel: surface Li abundance as a function of time 
($\log$ age) for the three
   models. The abundance is expressed as the ratio
 between Li and H numerical abundances ($N_{\mathrm{Li}}$/$N_{\mathrm H}$);
the starting of the MS phase and the turn off (TO)
 positions are indicated.
 Lower panel: temperature at the bottom of the convective envelope
   ($T_{\rm{cz}}$) vs. $\log$ age (when the star is
  fully convective the bottom of the convective core coincides with the center 
of the star);
 the indicative temperature of ignition for the $^7$Li+p
 reaction is $\sim2.5\times10^{6}$ K.
 }\label{Modelli}
 \end{figure}
 %==============================================
 %
 %============================================== FIGURA 2
 \begin{figure}
 \resizebox{0.5\textwidth}{!}{
 \includegraphics[width=3cm]{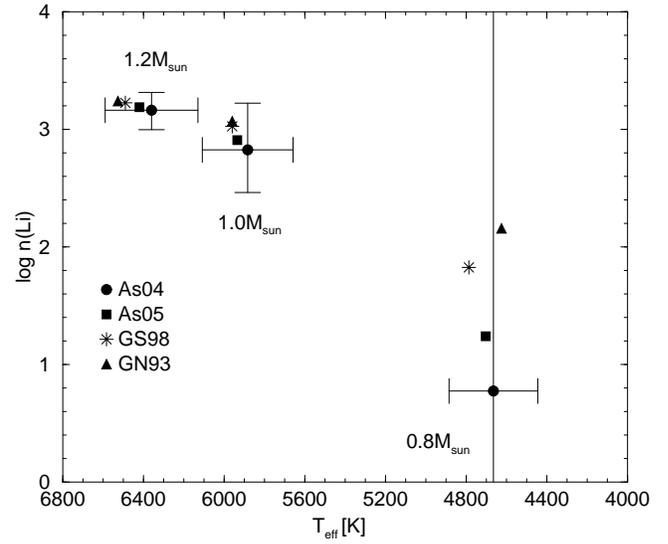}
 }
 \caption{Theoretical Li abundance 
(expressed as $\log n(\mathrm{Li})=12+\log(N_{\mathrm{Li}}/N_{\mathrm H})$)
 as a function of the effective temperature
 for 0.8, 1.0 and 1.2 M$_{\odot}$ at the age of $\sim$30 Myr. 
 Results for the four labeled different mixtures are shown
(Grevesse \& Noels 1993 -- GN93;
 Grevesse \& Sauval 1998 -- GS98; Asplund et al. 2004 -- As04, our
 ``standard'' mixture; Asplund et al. 2005 -- As05). For
 As04 we also show the error bars due to
 the estimated uncertainties on [Fe/H], $Y$ and $\alpha$ (see text). 
 }\label{LiTe}
 \end{figure}
 %==============================================

 %============================================== FIGURA 3
 \begin{figure}
 \includegraphics[width=7cm]{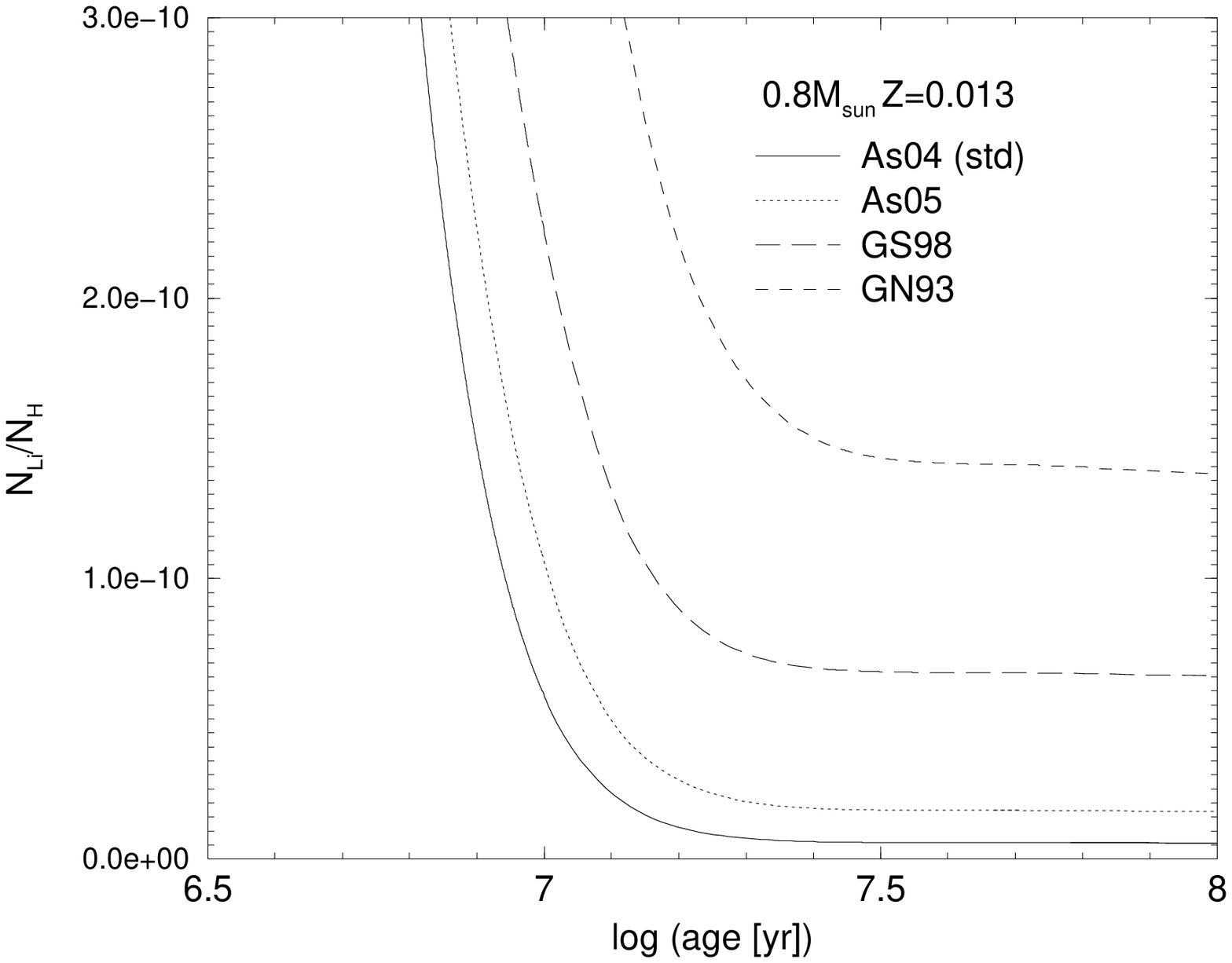}

 \includegraphics[width=7cm]{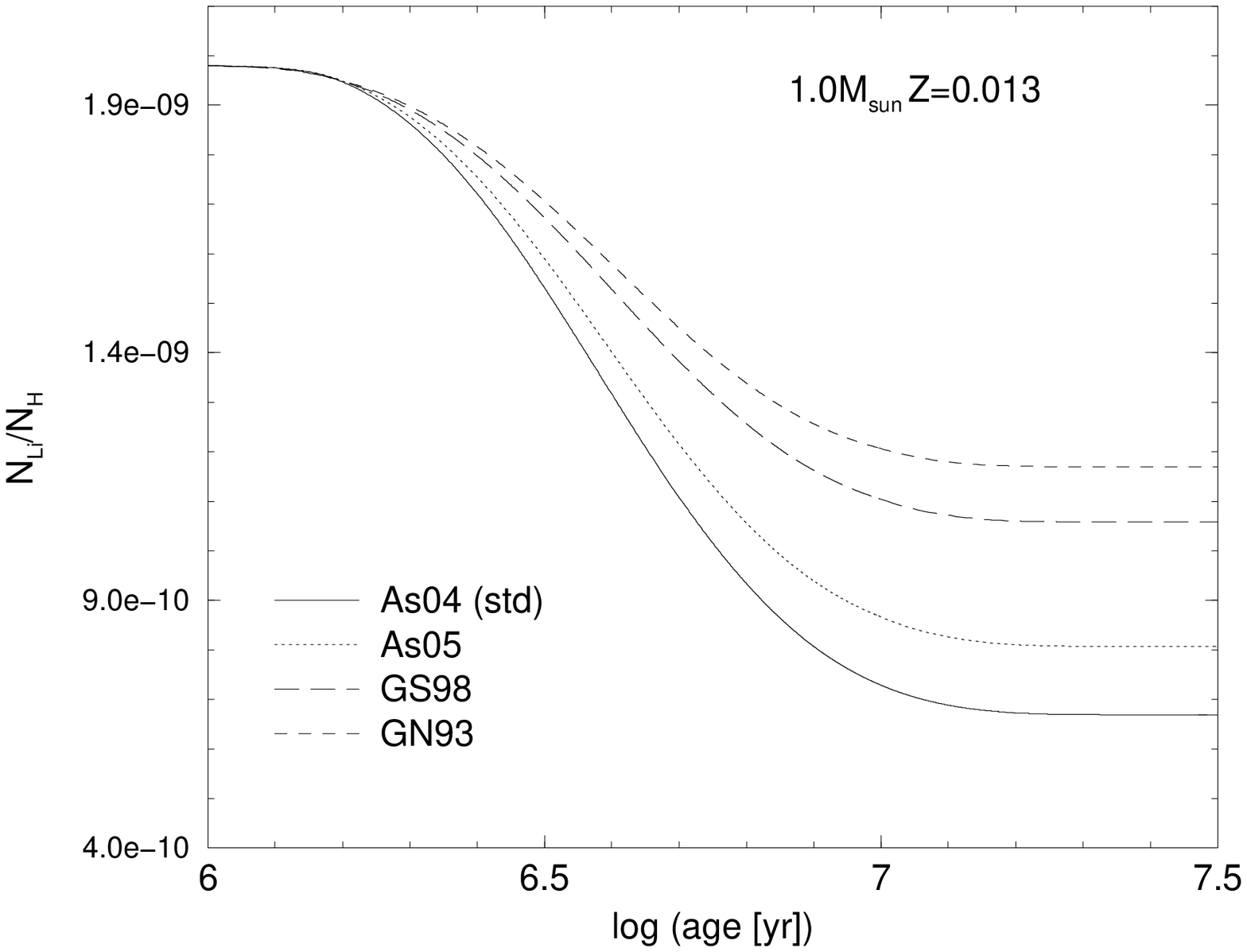}
 
 \includegraphics[width=7cm]{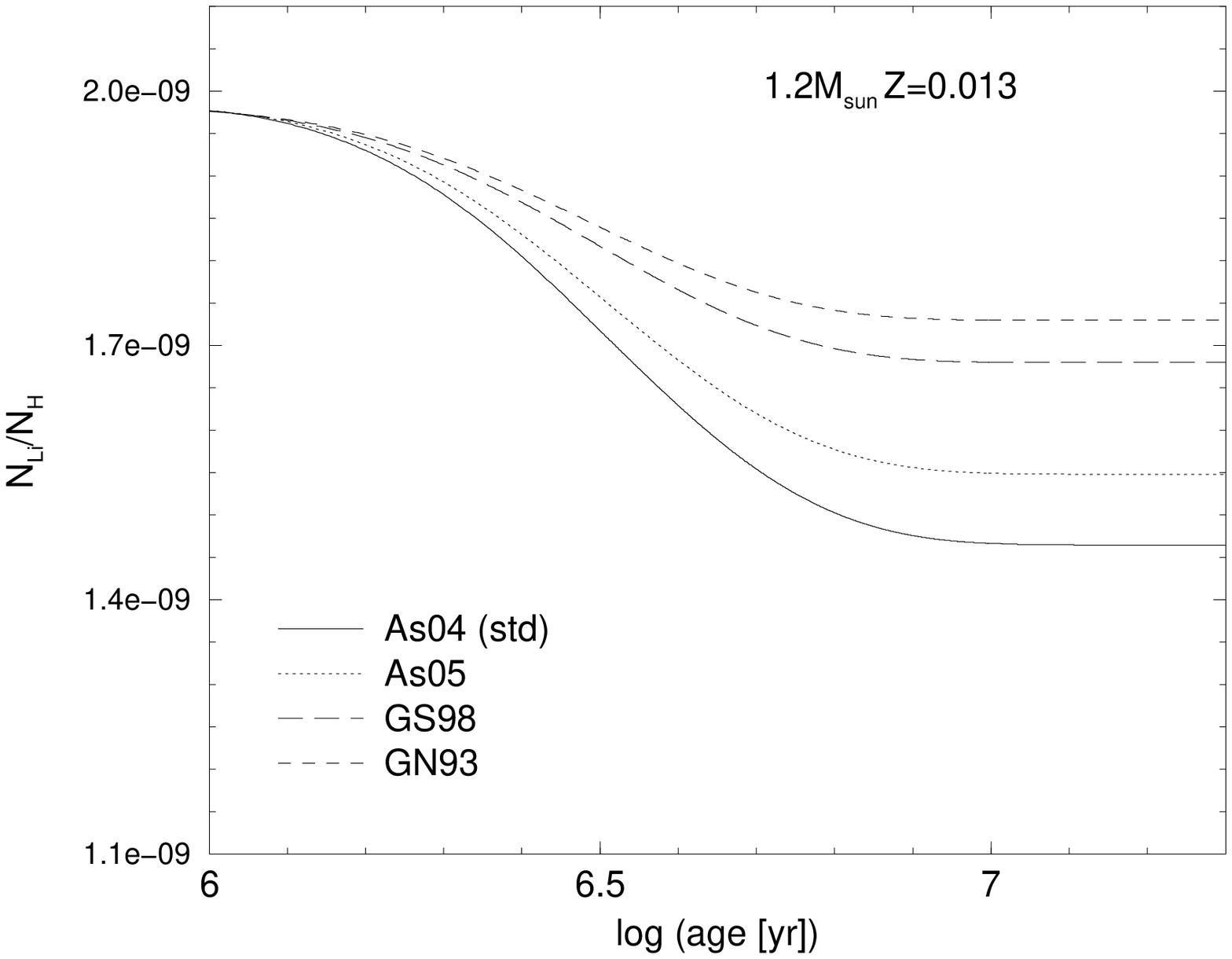}
 \caption{Surface Li abundance behavior during the
 PMS phase for the different labeled mixtures. The three selected masses
 are reported:
 0.8 $M_{\odot}$ (upper panel), 1.0 $M_{\odot}$ (middle panel), 
1.2 $M_{\odot}$
 (lower panel), for $Z$=0.013, $Y$=0.27, $\alpha$=1.9. 
Abundances and ages 
 are expressed as in Fig.~\ref{Modelli}; note the different scales for
Li abundances adopted in the three panels. 
 }\label{mixture}
 \end{figure}
 %==============================================
 %============================================== FIGURA 4
\begin{figure}
\resizebox{0.5\textwidth}{!}{
\includegraphics[width=7cm]{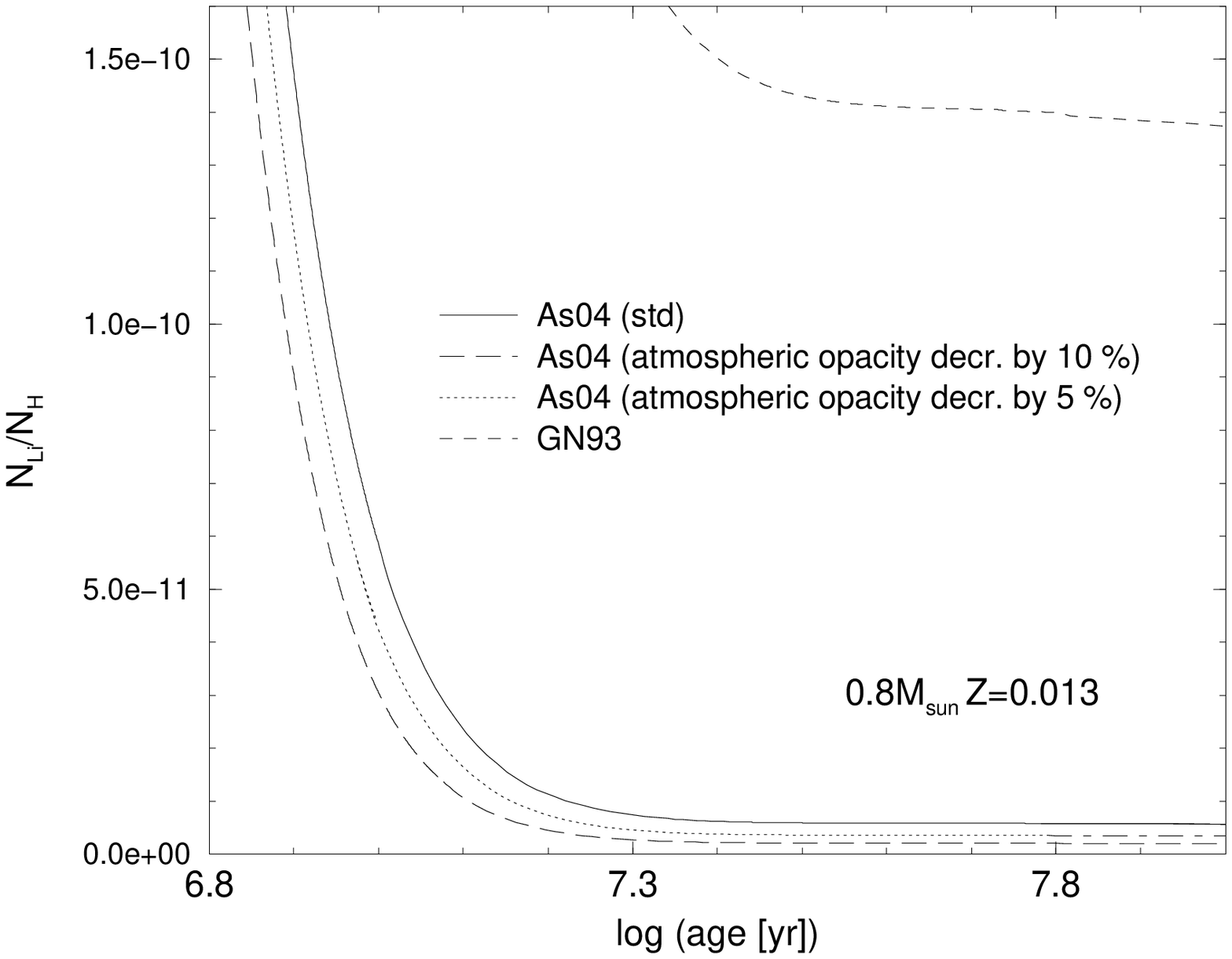}
}
\caption{Surface Li abundance vs.~age for
 0.8 $M_{\odot}$ ($Z$=0.013, $Y$=0.27, $\alpha$=1.9). 
The standard As04 model is compared to models
in which the atmospheric opacity has been decreased
by 5\% (dotted line) and by 10\% (long dashed line).
We report for reference also the model
with the mixture by GN93.
 }\label{opac}
 \end{figure}
 %==============================================
 
 \section{Li abundance for models with different
   mixtures}\label{risultati}

\subsection{Reference models}\label{risultati1}

We investigated three models of different masses (0.8, 1.0 and 1.2
$M_{\odot}$), representing 
the typical mass range covered by most observational data for Li
in open clusters; higher masses have very shallow convective envelopes and
thus the effect we are studying is too small to be accurately investigated.

The various models discussed in the present paper are
summarized in Table~\ref{tabella}:
Cols.~1 and 2 report the sequence number of the model
and the label used in the figures, respectively;
in Col.~3 the adopted solar mixture is given, while
Col.~4 lists the variations of elements with respect
to the solar mixture -- if any -- or the variations in the surface opacity or global metallicity.

 The upper panel of Fig.~\ref{Modelli} shows the evolutionary tracks for the
 three selected masses from the PMS to the bottom of the red giant branch
 (RGB) with our reference choice of metallicity and heavy element mixture
(model no.~1, As04, in Table~1). 
Note that for completeness we describe here the evolution up to the RGB;
however, the following discussions will focus only on
the evolution up to the ZAMS phase where
the largest amount of Li depletion occurs.
As
 well known, the treatment of the early PMS phases is very problematic and the
 definition of the birth line requires the adoption of hydrodynamical
 codes (see e.g. Stahler \& Palla \cite{stahler} for an extensive
 analysis). However, this is not a crucial point for our purposes
 since the stellar characteristic after the birth line are largely
 insensitive to the previous evolutionary history. Moreover, for the selected
 masses the equilibrium is reached very early (ages $\lesssim$ 1 Myr), when
they have not yet started surface Li depletion:
indeed, at the beginning of the PMS phase, the zone where the 
Li burning temperature is exceeded represents a small fraction 
of the total external convective region; the dilution is thus very large.
  For these reasons
 we will show our calculations starting from an age of 1 Myr.

 The middle panel of Fig.~\ref{Modelli} shows the behavior of Li
 abundance ($N_{\rm{Li}}$/$N_{\rm{H}}$) as a function of the age
 for the same models, while in the lower panel the temperature at
 the bottom of the convective zone ($T_{\rm{cz}}$) is reported. 
As well known, the smaller
 the mass the deeper the convective zone and thus the higher  
 the temperature at its lower edge.
 During the PMS the extension of the convective envelope and $T_{\rm{cz}}$ decrease
 until, in the cases of 1 and 1.2 $M_{\odot}$, it becomes lower than $\sim2.5\times10^{6}$ K;
 as a consequence Li destruction stops
 and the surface abundance decreases
 only due to microscopic diffusion. On the contrary, for the model with 0.8
 $M_{\odot}$ Li burning is efficient also during the MS phase 
even if the largest amount of Li is depleted
 during the PMS phase (note that MS Li depletion
is not evident from the figure due to the large scale adopted:
for instance, from the ZAMS to the turn off,
$N_{\rm{Li}}$/$N_{\rm{H}}$ decreases from $\sim{6\times{10^{-12}}}$ to $\sim{1.5\times{10^{-13}}}$, i.e. by a factor $\sim$40).
 
 After central hydrogen exhaustion, the H burning continues in a
 shell surrounding the He core and the stars move to lower effective
 temperatures. During this phase the convective envelope extends into
 regions in which Li has accumulated due to the
 effect of diffusion, and the external abundance temporarily increases;
 then, it decreases below the level of detection when the external 
convective region
 reaches the layers in which Li has been completely burned (dilution effect).

The details of Li depletion depend on
 the adopted physical inputs, in particular the radiative opacity and
 chemical composition for a given mixing scheme.  Thus, the situation shown in
 Fig.~\ref{Modelli} has to be considered as a description of the
 dependence of Li depletion on the mass, while the detailed quantitative
 values hold only for the labeled parameters.

\subsection{Global uncertainties}\label{risultati2}

Empirical data and
 theoretical predictions are
 usually compared in the [\nli, \teff] plane; thus it is
 important to evaluate the ability of canonical models to reproduce these
 quantities. The present 
empirical uncertainties on the 
 metallicity, He abundance and efficiency of the external convection 
 (usually  parameterized by the mixing length value $\alpha$) 
play an important
 role in the comparison between theory and observations. 

At the typical effective temperatures of the stars under investigation
the surface He abundance
cannot be measured spectroscopically.
 As a consequence one has to rely on indirect methods, such as assuming a
 relation between metal and He enrichment of the Galactic medium which
 is not well defined (see, e.g., Pagel \& Portinari \cite{pagel};
 Castellani, Degl'Innocenti \& Marconi \cite{caste99}; Luridiana et al. \cite{luridiana}; Olive \&
 Skillman \cite{olive}). 
Under this assumption, $\Delta{Y}$=$\pm$0.02 is generally assumed as 
an estimate of the
 original He abundance uncertainty for these stars.  

Regarding  the Fe content, 
disk stars cover a wide interval of [Fe/H] 
and typical uncertainties depend on several factors such as
on whether the analysis is spectroscopic or photometric,
the quality of the data, the method of analysis, statistics etc.
In the case of young nearby stars
(see e.g. Chen et al.~\cite{chen};
Bensby et al.~\cite{bensby})
and open clusters with nearly solar metallicity
(e.g. Sestito et al.~\cite{sestito03};
Carretta et al.~\cite{carretta04}; Randich et al.~\cite{R06})
the quoted errors range from $\Delta$[Fe/H]$\sim$$\pm$0.01
up to $\sim$$\pm$0.06$\div$0.07, when
the analysis is based on a large sample of stars observed with
high resolution spectrographs. Errors can be as high as
$\Delta$[Fe/H]${\sim}\pm$0.1 (or  more) if
results from low resolution spectroscopy or from the photometry
are considered (see e.g. Friel et al.~\cite{friel02};
Nordstr\"om et al.~\cite{nord}). 
We adopt here a value 
of $\Delta$[Fe/H]=$\pm$0.1 as a conservative estimate of
observational uncertainties on the metallicity of solar-type stars.

Finally, the $\alpha$ parameter adopted in the calculations, depending on
the selected set of atmospheric models, could be calibrated by requiring the
reproduction of the color of the MS stars of the observed open
cluster. However $\alpha$ is a useful parameter to take into account our
lack of precise knowledge about the convection mechanisms, and there is no
reason for which it must assume the same value for stars of different mass
and chemical composition, or even for different evolutionary phases of 
the same
star. When a direct calibration of the $\alpha$ value for PMS
stars through the reproduction of their stellar colors is not possible, one
must take into account an uncertainty on this parameter;
we allow a variation of the mixing length parameter $\alpha$ by
$\pm$0.3. 

 We computed six models for each of the three selected masses by fixing, in
 turn, one of the three parameters at its boundary values (lower and upper)
 and keeping the others at their central values ($Z$=0.013, $Y$=0.27,
 $\alpha$=1.9). In this way,
we obtained relative variations of
 Li abundance and effective temperature with respect to our ``standard''
 model, representing
 an evaluation of the uncertainties on Li and \teff~ due to plausible
 empirical errors on $Z$, $Y$ and $\alpha$. Then, 
the relative variations of
 Li and \teff~ (on the increase and on the decrease) due to each of the
 quoted parameters are added quadratically to give a quantitative estimate of
 the total (upper and lower) uncertainties.

In Fig.~\ref{LiTe} the results for our ``standard'' models for the three
different masses at 30 Myr (when for all the
stars 
the main Li depletion has ended) are
shown in the [\nli, \teff] plane with the error
bars due to the ranges of [Fe/H], $Y$ and $\alpha$, calculated as
described above. 
The error bars
represent the range of predicted theoretical values
due to the uncertainties discussed above, while intrinsic
errors of the model (due to physical inputs) are not considered;
therefore the error bars are an indicative
estimate of the uncertainties in the initial physical parameters. 
Note that among the three quoted error
sources, the largest
contribution to the total error bar is due to the uncertainty on $Z$; 
in particular, the effect of 
the metallicity variation is very strong for 
the lowest mass, for which 
\nli~appears to be unpredictable.
 
\subsection{Heavy element mixture variations and opacity effect}\label{risultati3}

The next step is aimed to evaluate the effect 
of a
variation of the adopted heavy element mixture with respect to the 
sources of uncertainty described above. Thus we calculated models for the 
same ($Z$, $Y$, $\alpha$) and 
the Grevesse \& Noels (1993),
Grevesse \& Sauval (1998) and Asplund et al.~(2005) mixtures
(models no.~2, 3, 4 in Table~1; GN93, GS98, As05).  Since the
temperatures of main interest for Li depletion are those at the bottom of the
external convective region during PMS Li burning, that is
$T_{\rm{cz}}\sim3\div4\times10^{6}$ K (see Fig.~\ref{Modelli}), we calculated
high temperature opacities (T$>$ 12000 K) for the chosen mixtures using
the procedures available on-line from the OPAL
group.
%\footnote{http://www-phys.llnl.gov/Research/OPAL/new.html}. 

The results of our calculations are reported in
Fig.~\ref{LiTe}.  The variations in Li
abundance and \teff~ due to the mixture changes are within the
(conservative) previously quoted uncertainties but they are not
negligible. It can also be noticed that
the smaller the mass, the more 
sensitive is the dependence of the Li abundance on
the adopted physical parameters.

For completeness Fig.~\ref{mixture} shows the surface Li abundance behavior
during the PMS phase for the three selected masses with the four different
mixtures of Fig.~\ref{LiTe}. The models that undergo
the largest amount of Li depletion
have higher opacity in the region at the bottom of the
convective zone, and thus deeper and hotter bottom convective envelopes
(see Sect.~\ref{discussione}).

 However, the various quoted mixtures differ in the
 abundance of many elements.  In the following
(Sect.~\ref{discussione}), we will thus discuss simpler
 situations in which only one or a few critical element abundances are changed.  This
 provides a better understanding of the dependence of the opacity at the
 bottom of the convective envelope -- and thus of Li depletion -- on the
 element abundance variations.

Low temperature (i.e. atmospheric) opacities calculated for the 
As04 and As05 mixtures are
not available; however, since the temperature of the bottom of
the convective zone is
relatively high, the effect of a change in
the surface opacity due to the mixture update 
on Li depletion should not be
too relevant. We expect that changes in atmospheric opacities
mostly affect the model with the
smallest mass (0.8 $M_{\odot}$), which has the lowest effective temperature
and the deepest convective envelope. From the calculations by the OPAL group
we found that, for a 0.8 $M_{\odot}$ star and at a temperature of about 12000 K (the lowest
$T$ for which the OPAL tables are computed),
the change in the internal opacity values when going from the
GN93 to the As04 mixture (at fixed $Z$) is less than 5\%;
we took this value as representative also of a possible variation
in the lower temperature opacities. 
In order to analyse the effect of such a variation,
we evolved a 0.8 $M_{\odot}$ ($Z$=0.013, $Y$=0.27, $\alpha$=1.9) 
with the usual atmospheric opacities (Alexander \& Ferguson 1994) 
artificially decreased by 5 and 10\%
(models no.~5 and 6 in Table~1). For the interior opacities we adopted 
 the As04 mixture. The results for Li
depletion are shown in Fig.~\ref{opac} compared with the
``standard" 0.8 $M_{\odot}$. The effect is rather small; 
the decreased atmospheric
opacities result in models with hotter external regions,
thus, even if the depth of the convective envelope is nearly unchanged, their
bottom temperature is slightly higher and 
a slightly larger amount of Li depletion
occurs. Nevertheless, 
the effect is negligible with respect to
that due to a mixture variation for the interior:
the ZAMS Li abundances (age $\sim$ 30 Myr)
decrease by factors $\sim$1.5 and 3 (with respect to
the As04 standard  model) when surface opacities
are decreased by 5 and 10 \%, respectively; on the other hand,
going from  the standard As04 mixture 
to the GN93 one, the ZAMS Li content
increases by a factor $\sim$18.

 \section{Interpretation of the Li abundance behavior}\label{discussione}
  
 In order to understand the effect of the different mixtures 
 on the extension of the convective envelope, 
we analyse the contribution of the different elements
 to the radiative opacity at the temperatures of interest, 
 i.e. those at the base of the convective envelope during the PMS.  
Therefore, we computed opacity tables for pure elements by 
using the data and routines 
 available on-line form the LAOL group\footnote{http://www.t4.lanl.gov/cgi-bin/opacity/tops.pl} (we
did not use OPAL opacities in this case
because they are not available for a single element).

\begin{figure}
 \includegraphics[width=7cm]{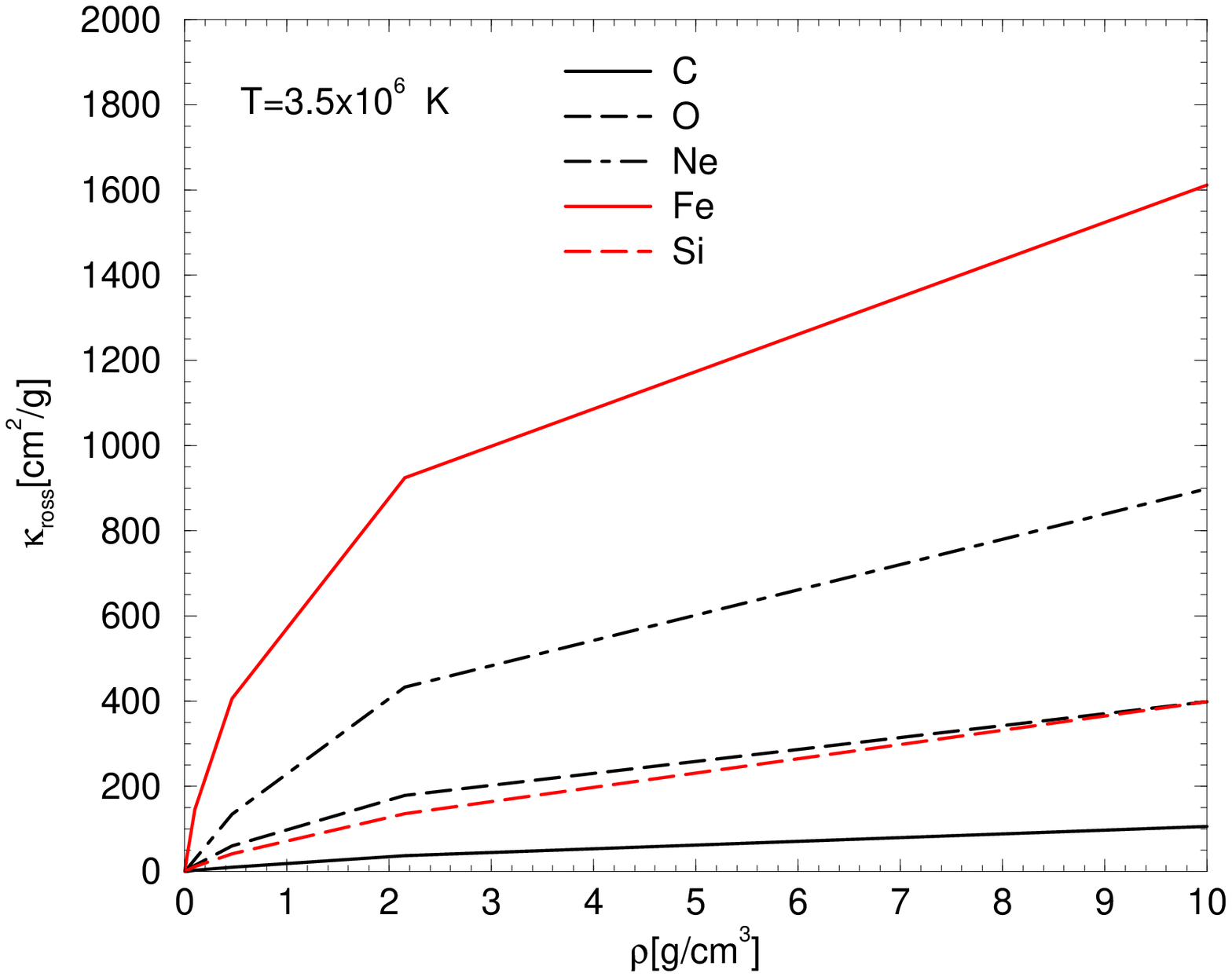}
 \includegraphics[width=7cm]{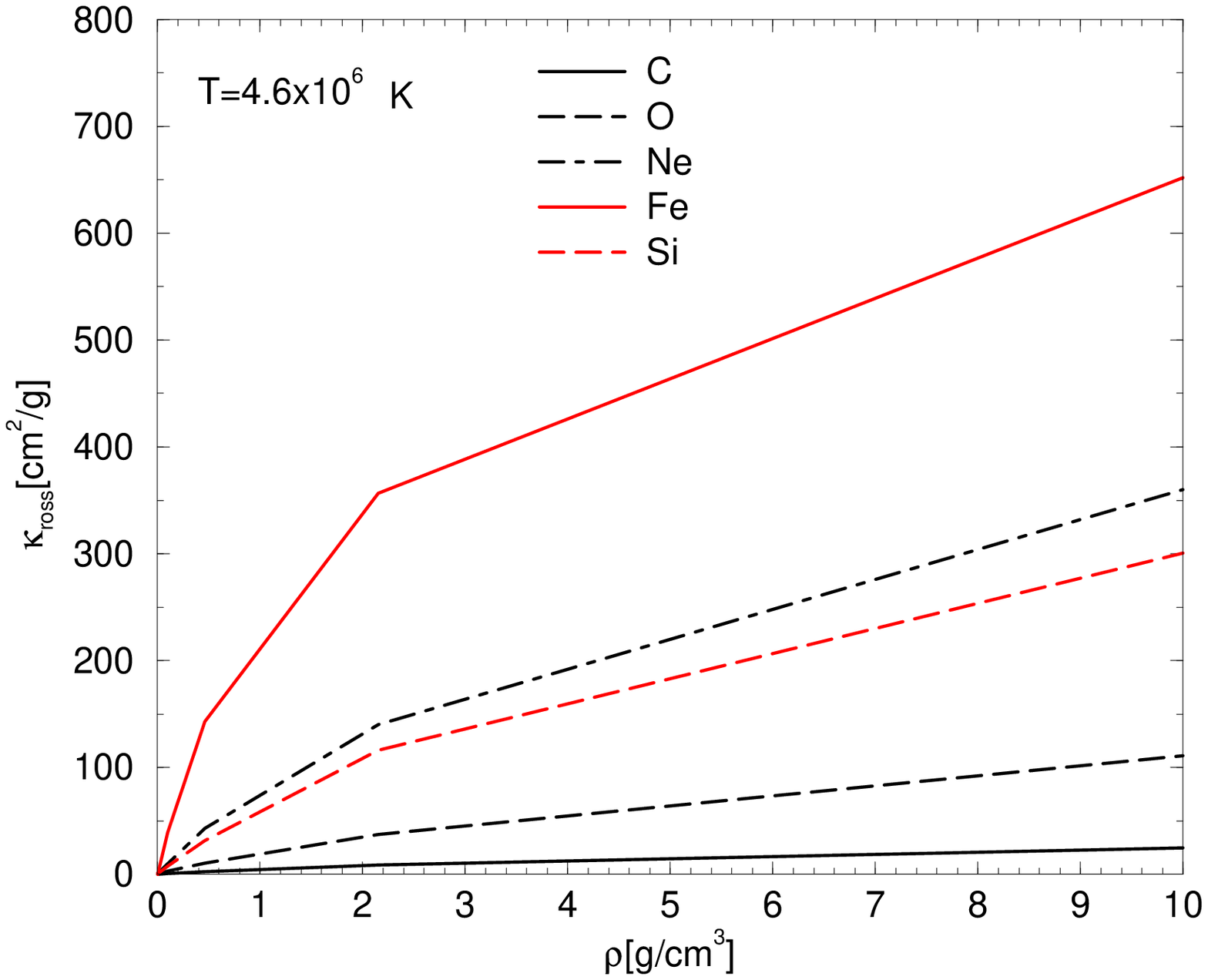}
 \caption{The Rosseland opacity for $T=3.5\cdot10^{6}$ K (left panel) 
 and $T=4.6\cdot10^{6}$ K (right panel) as a function of the density
 for several pure elements, as labeled. The opacity values are obtained from the
 URL http://www.t4.lanl.gov/cgi-bin/opacity/tops.pl.}\label{elementipuri}
 \end{figure}

 Figure \ref{elementipuri} shows the pure element opacity for the most
 important elements: H, He, C, O, Ne, Fe, Si.  The opacities are calculated
 for $T=3.5\cdot10^6$ K and $T=4.6\cdot10^6$ K 
 and for densities ranging from
 0 to 10 g cm$^{-3}$.  Note that at the selected temperatures the most opaque
 element is Fe, followed by Ne, and O is more opaque than C; the
 opacity of pure He and H is negligible with respect to those of metals.
 Obviously in a realistic case these single element opacities must be weighted with
 the abundance of each element in a mixture in which H and He are present too.

 The next step to understand the results of Sect.~\ref{risultati}
  is thus to compute PMS models
 in which, for the standard metal and He abundances,
the contribution to the metallicity is made by only one element.
Therefore, we evolved 
the models for $Z$=0.013 and $Y$=0.27 by assuming that the
contribution to the metallicity is made by only C, O or Ne;
we also took into account the variation of the composition in
  calculating the nuclear reaction rates, while for the other thermodynamical quantities
  we adopted the OPAL EOS\_2001 tables calculated for the Grevesse (1991)
  solar mixture.
Our assumptions, as a first
approximation, affect the PMS evolution only through the variation of the
stellar opacity; we checked that the temperature profile in the convective
envelope does not change in a significant way for the three models.

\begin{figure}
 \includegraphics[width=7cm]{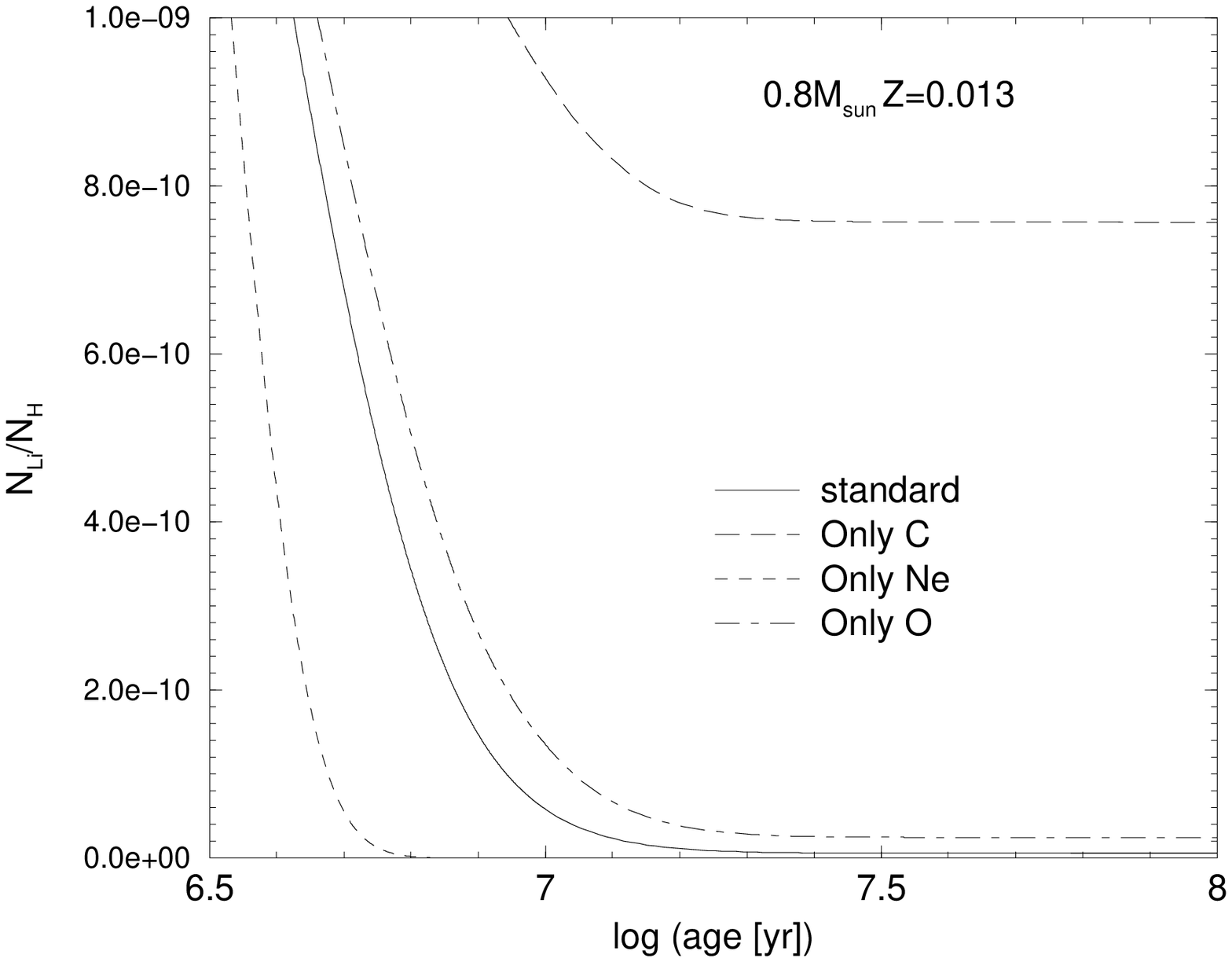}

 \includegraphics[width=7cm]{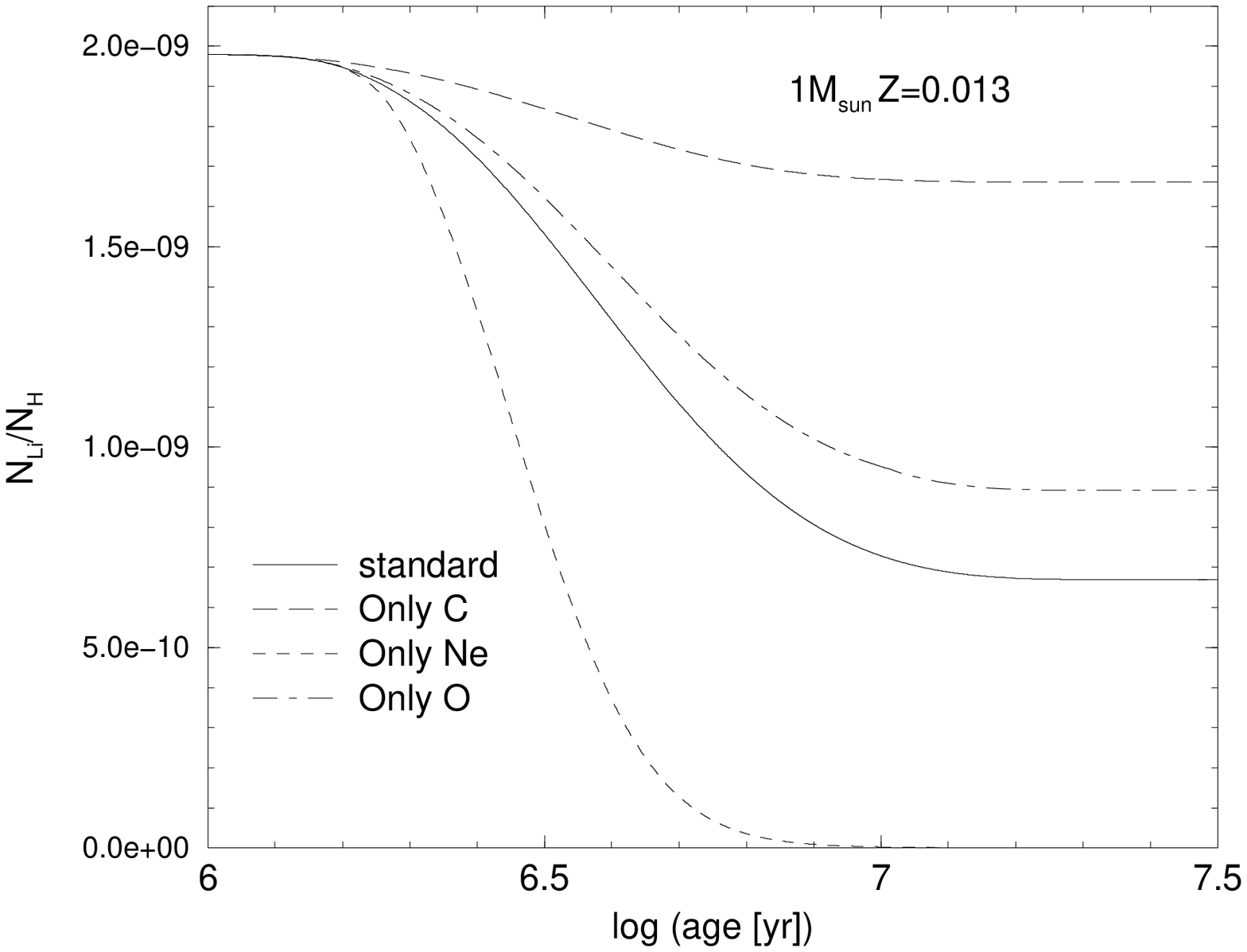}
 
 \includegraphics[width=7cm]{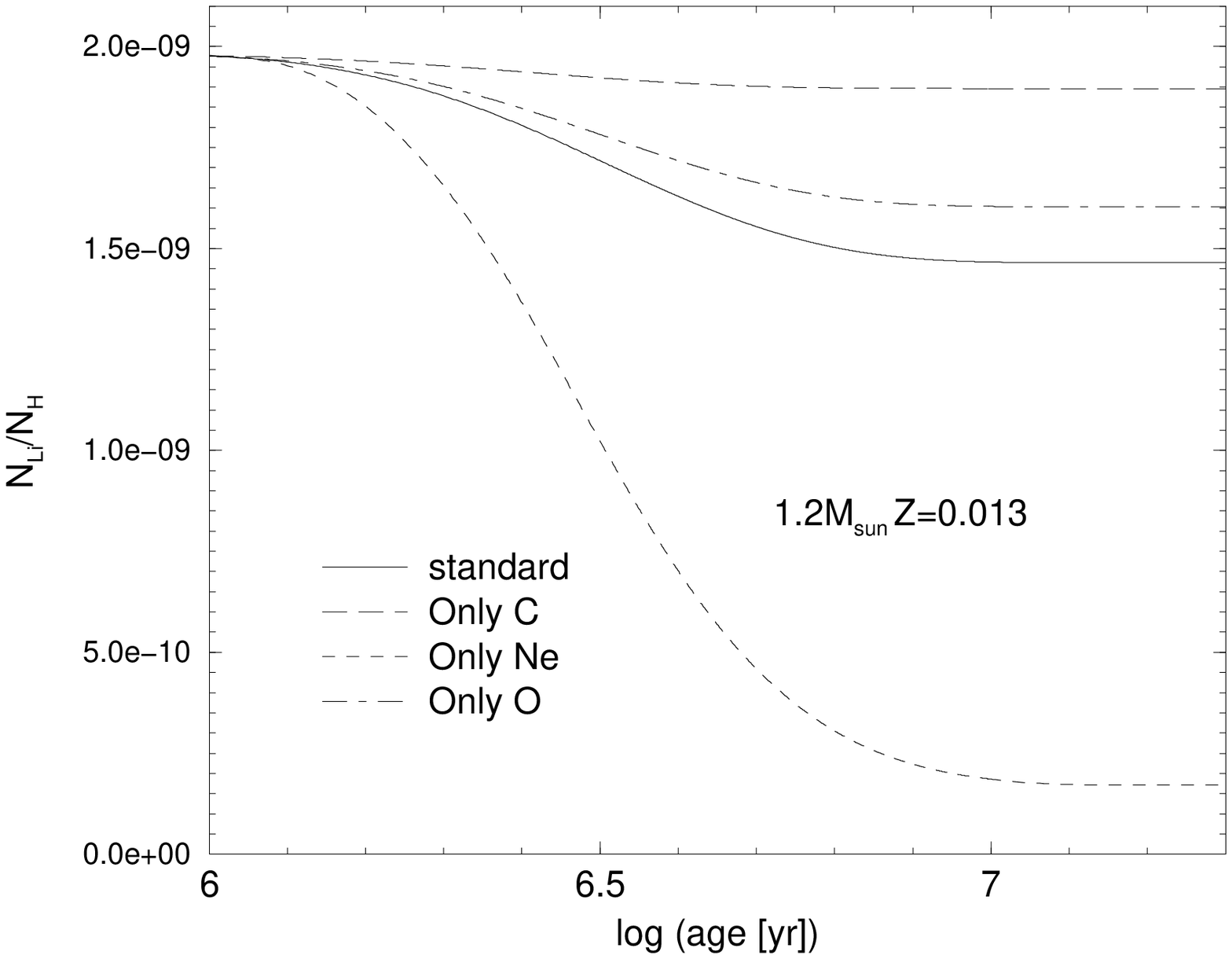}
 \caption{Surface Li abundance as a function of time for PMS models
 of 0.8 (upper panel), 1.0 (middle panel) and 1.2 (lower panel) $M_{\odot}$.
 For each mass Li abundance is calculated for $Z$=0.013,
$Y$=0.27, $\alpha$=1.9 for
 a metallicity due to only one element: C, O, and
 Ne. The results are compared with the ``standard''
 model, where the As04 mixture is adopted. Note the different scales
adopted in the three panels.
  }\label{figsolo}
 \end{figure}

The opacity tables were computed by adopting the procedures 
 available on-line from  the OPAL group. The results of these numerical experiments are
 shown in Fig. \ref{figsolo} (models no.~7, 8, 9 of Table~1); the behavior of
 Li abundance is easily explained by the opacity values seen in
 Fig. \ref{elementipuri}: the element with the highest opacity at temperatures
 and densities typical of the base of the convective envelope leads to the
 highest Li depletion for the models of Fig. \ref{figsolo}.  In addition, the
 results for our ``standard'' models are not so different from those obtained
 for a metallicity contributed 
by only O: this is easily understandable, since O is
 the most abundant element (after H and He) and its opacity is greater than
 that of C (which is the next most abundant element). The behavior is similar
 for the three masses although, as already noted in Fig.~\ref{mixture}, the
 effect is smaller for larger stellar masses.
 
 %============================================== FIGURA 5
 %============================================== 
 %============================================== FIGURA 6
 %============================================== 
 %============================================== FIGURA 7
 \begin{figure}
 \includegraphics[width=7cm]{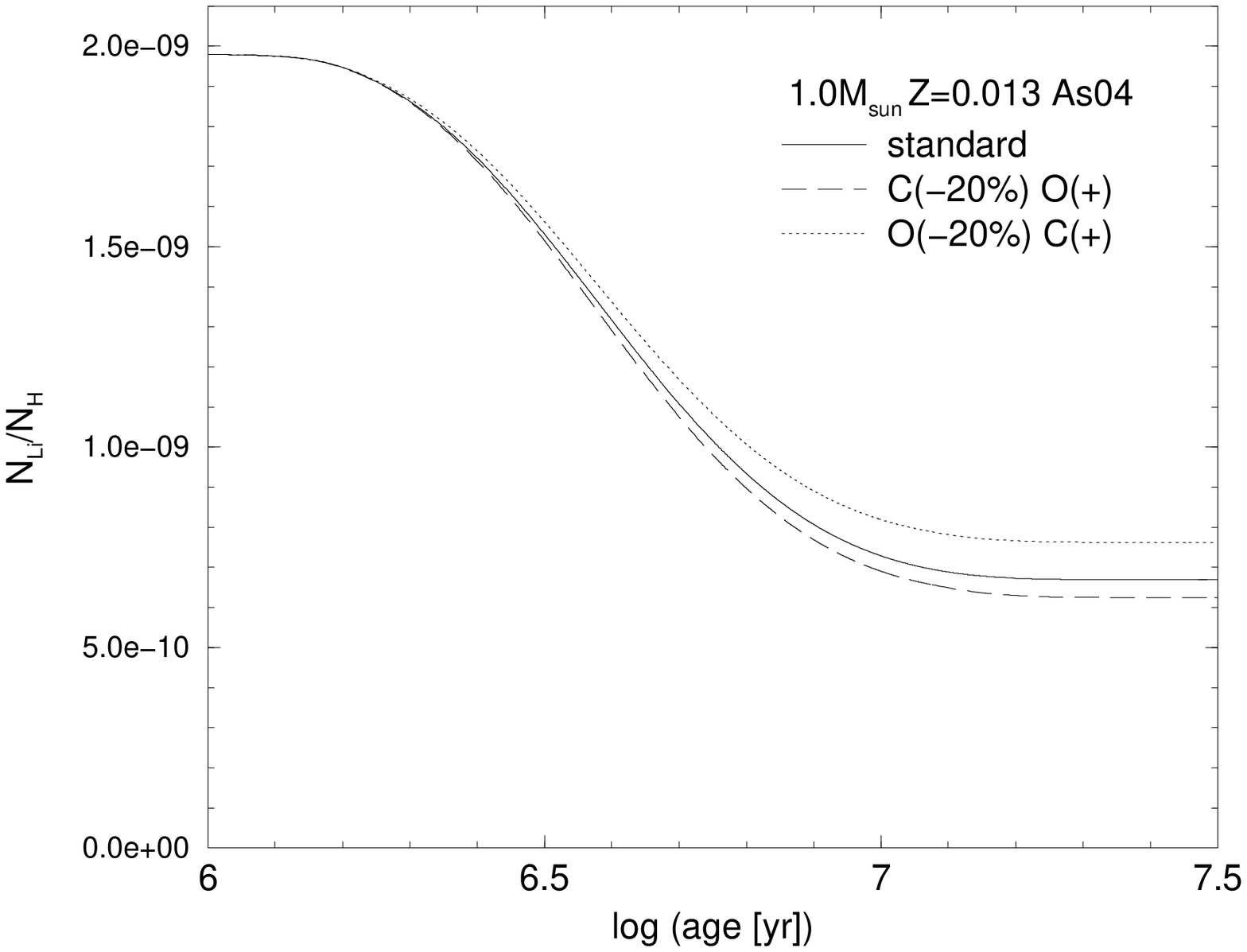}

 \includegraphics[width=7cm]{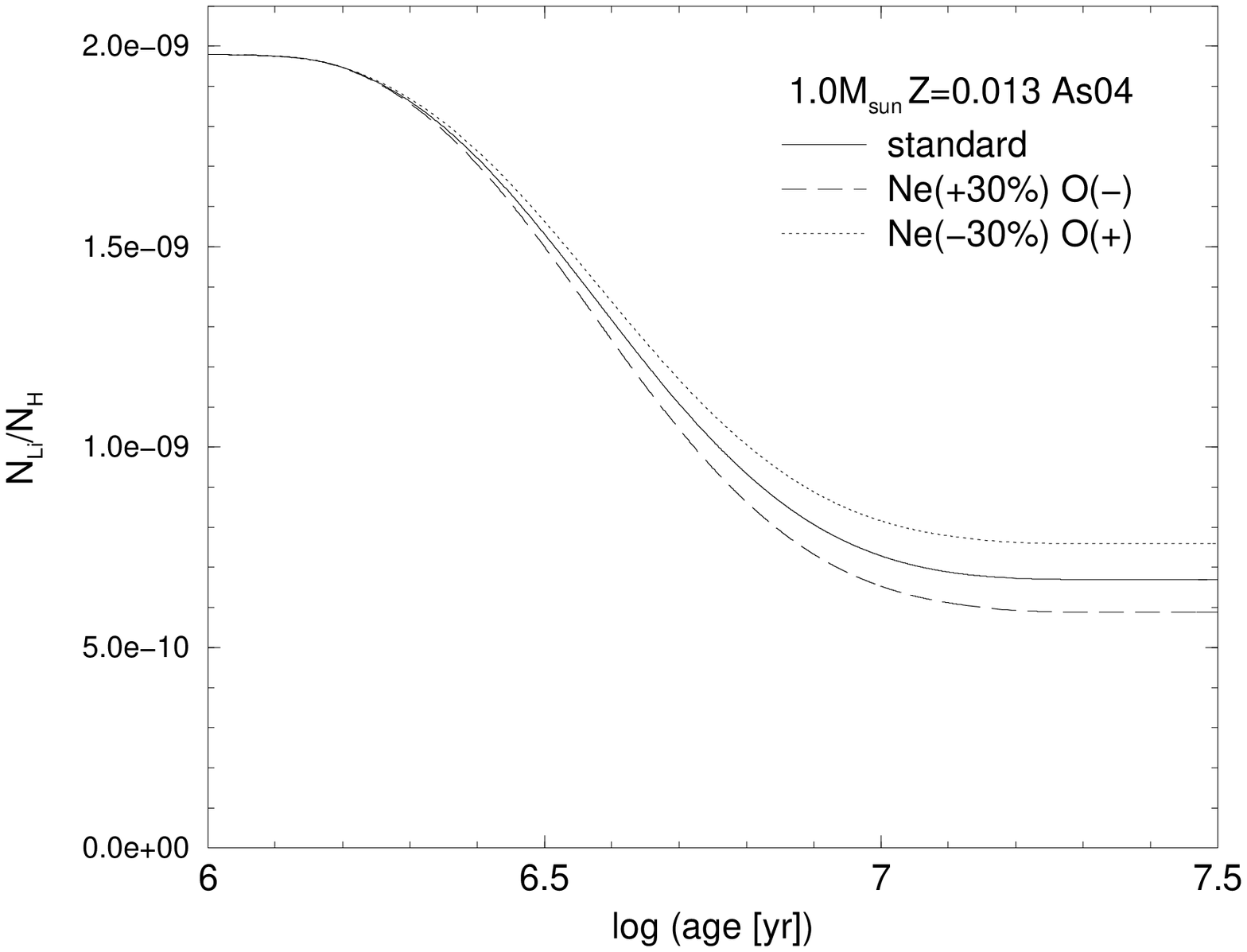}

 \includegraphics[width=7cm]{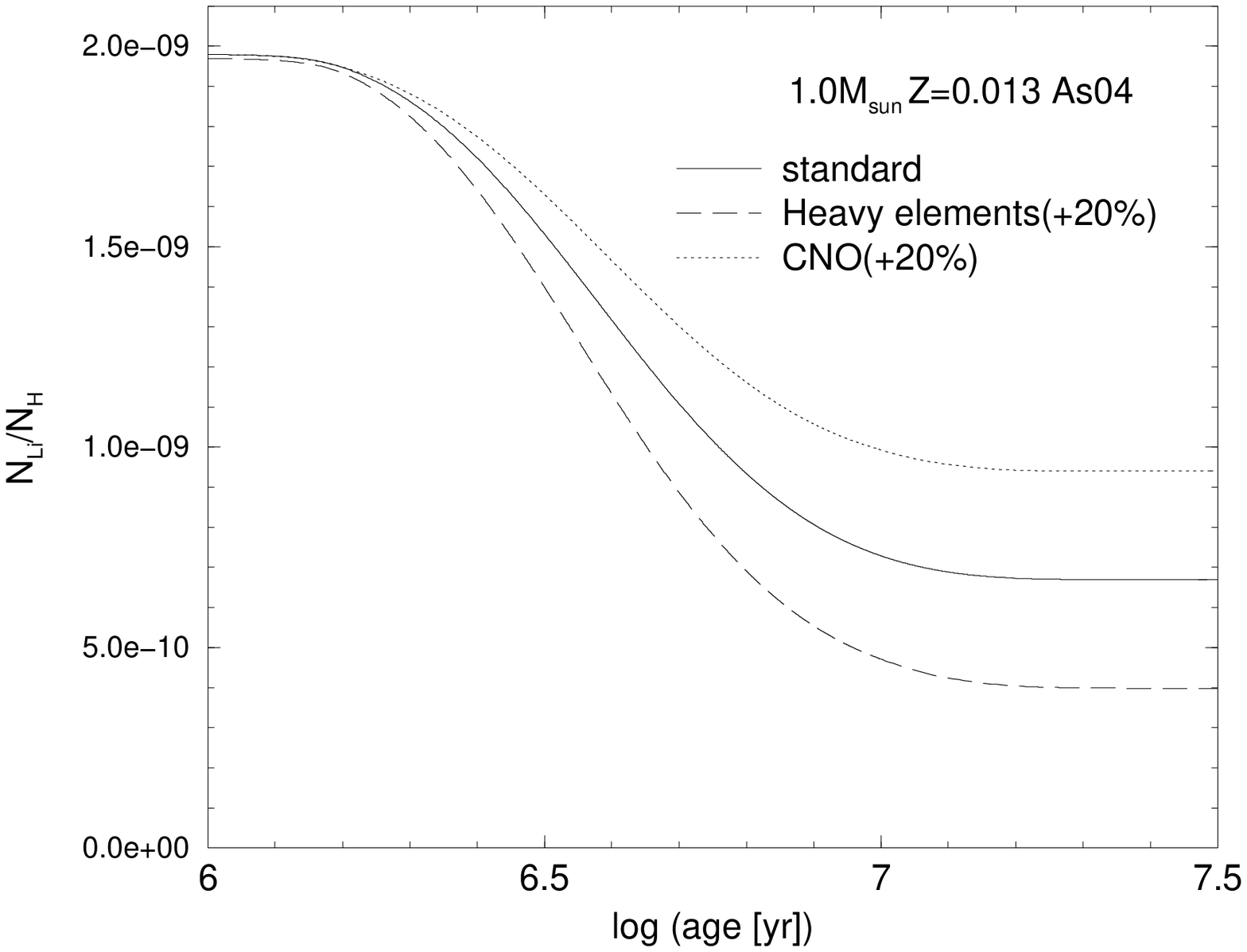}
 \caption{Li abundance vs.~time for PMS models of 1 $M_{\odot}$.
We adopted, as usual, $Z$=0.013, $Y$=0.27, and $\alpha$=1.9
and we varied the mass fraction of some elements starting from the As04
mixture
(see text).
 }\label{figrisul}
 \end{figure}
 %============================================== 

A more realistic situation is
represented by models with our ``standard'' (As04) mixture in which only few
elements are varied each time by $\sim20\div30 \%$, keeping the
total metallicity constant. This could
reflect not only an update of the solar mixture, but also a possible
scatter in the composition of an open cluster:
indeed,
although stars in the same cluster are usually assumed to have the
same chemical composition, an intrinsic dispersion in [Fe/H]
(up to $\sim\pm$0.15 dex) among cluster members cannot be excluded
(e.g. Nissen~\cite{nissen}; Schuler et al.~\cite{schuler05}). Analogous
variations may exist for other elements. In Fig.~\ref{figrisul} we show only
the results for 1.0 $M_{\odot}$, since the overall behavior was the same as in
the previous calculations.  The upper panel of Fig.~\ref{figrisul} shows the
effects on Li depletion when C and O are changed maintaining the sum
unaltered in mass
of the two elements ($X_{\mathrm C}$+$X_{\mathrm O}$).  In the first case C is
decreased by 20\% in mass while in the second case O is decreased by the same
amount (models no.~10 and 11 in Table~1). An enhancement in O abundance
increases the Li depletion, due to the higher opacity at the bottom of the
convective zone and thus to the deeper extension of the convective envelope;
the opposite happens when O is decreased.  However, the effect is not
symmetric: this is due to
the fact that we keep the sum of the abundance in mass of the two elements -- and thus the 
total metallicity -- constant, and
the two elements have  different
atomic weights and ``standard'' abundances; therefore,
C and O are varied by different total amounts 
in number. These variations
have 
different effects on the global mixture and on the opacity, and thus
on Li depletion. The middle panel of Fig.~\ref{figrisul} shows the
results of a decrease/increase by 30\% in mass of the Ne mass fraction (with a
corresponding O variation to preserve $X_{\mathrm O}$+$X_{\mathrm{Ne}}$;
models no.~12 and 13 in Table~1). Due to the higher Ne opacity,
an increase/decrease of Ne abundance leads to a decrease/increase of the Li
surface abundance but, in this case, the effect is obviously symmetric.  
The last numerical
experiment (bottom panel of Fig.~\ref{figrisul}) evaluates the effects of
increasing/decreasing the relative abundance of CNO elements and elements
heavier than O (models no.~14 and 15 of Table~1): an increase in CNO with
respect to heavier elements produces a lower depletion,
and vice versa.
The results of all the previous numerical experiments can be understood 
in terms of the radiative opacity of the heavy elements. In particular,
from the point of view of Li depletion of low mass stars, 
the maximum depth reached by the convective envelope is very important, 
which in turn is
sensitive to the opacity in the temperature range 3$\div$4$\cdot$10$^6$ K
(see Fig.~\ref{Modelli}). In
this interval the most opaque elements, among those with a significant
abundance, are the elements
heavier than O. If the abundance of these
elements is higher, then
the mixture is more opaque and thus
the convective envelope is deeper.  At fixed metallicity ($Z$=0.013), the total
abundance of the elements heavier than O increases progressively in the
following mixtures: GN93, GS98, As05 and As04, thus explaining
the behavior shown in Fig.~\ref{mixture}.

%\section{Cluster metallicity from the observed [Fe/H]}
\section{Models with fixed [Fe/H]}\label{fixedFe}

In the previous sections we analysed the effects of a 
mixture variation at fixed 
global metallicity: however, the variation in the solar 
composition also affects the conversion of 
the [Fe/H] value of the observed stars (derived through a spectroscopic
analysis) to 
the total metallicity $Z$.
This point is very subtle because to compute stellar models one needs
both the Fe abundance and the total metallicity $Z$; the two
quantities are related through the distribution of elemental abundances. In
general, unless individual abundances are derived, a solar mixture is adopted. As a
consequence, a revision of the photospheric solar abundance directly leads 
to a variation of the inferred total metallicity from the observed [Fe/H].
For solar-like stars, [Fe/H] can be estimated by means of differential
analysis,
that is by comparison of the Fe lines of the star with those of the Sun. In
such a case, the numerical value of [Fe/H] is unaffected by a change of the
solar mixture and the updated global metallicity of the star can be easily
derived by adopting the new value of $(Z/X)_{\odot}$. That is:
 
$$
\log (Z/X)_*= \rm{[Fe/H]} +  log (Z/X)_{\odot}.
$$

\begin{figure}
\resizebox{0.5\textwidth}{!}{
 \includegraphics[width=7cm]{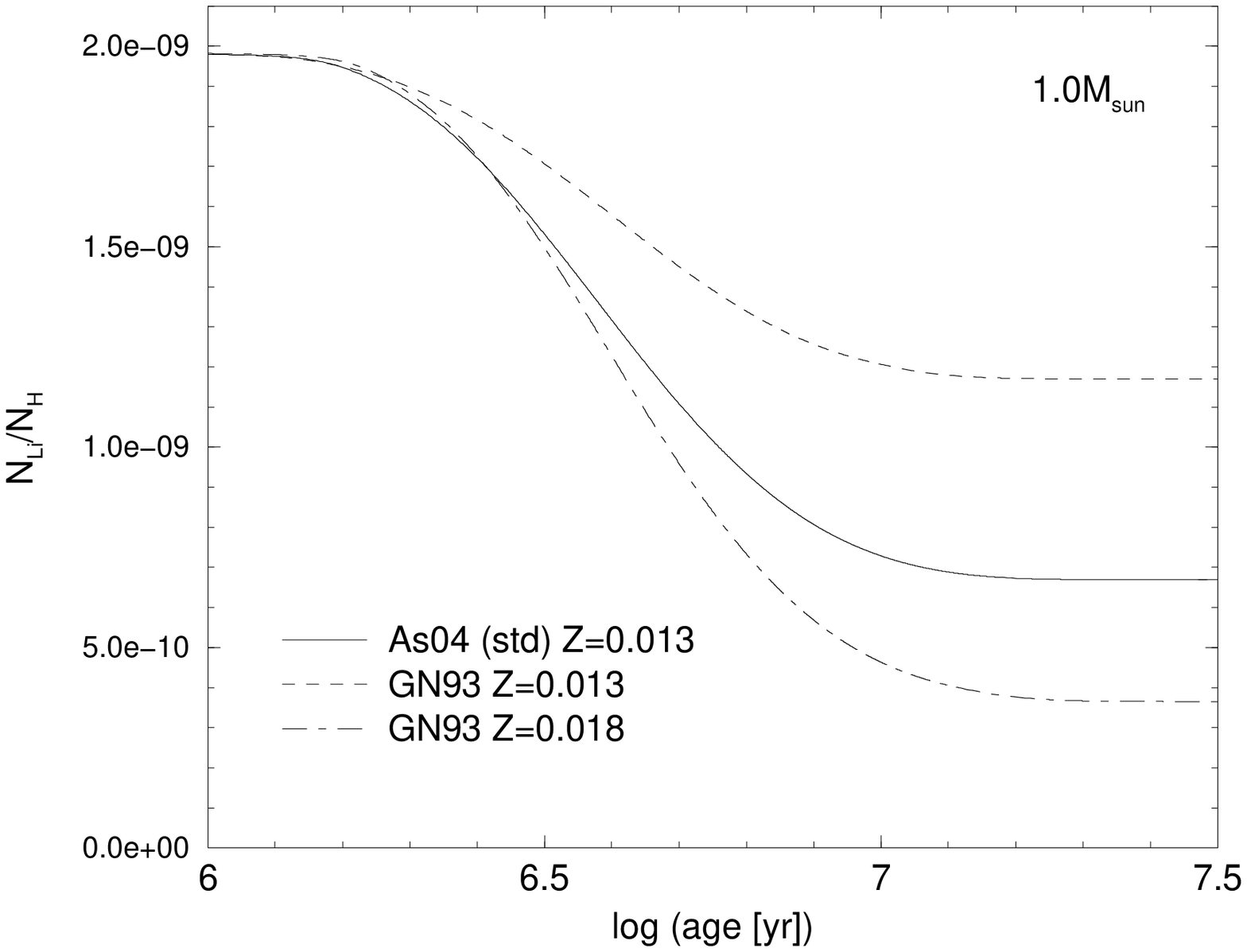}
}
 \caption{Surface Li abundance vs.~age for 
our standard model (As04 mixture with $Z=0.013$) compared to a model 
with $Z=0.013$ and GN93 mixture (the same of Figs.~\ref{LiTe},
\ref{mixture} and \ref{opac}) and  
to a model with the GN93 solar composition and metallicity $Z=0.018$
(see text).}
\label{Zchanged}
 \end{figure}

In order to give a quantitative estimate of the change of $Z$
(at fixed [Fe/H])
when the mixture and $Z/X_{\odot}$ are altered, 
for the
GN93 composition [Fe/H]=0 corresponds to Z$\sim$0.018, while for the As04
mixture, 
as already discussed, the corresponding metallicity is
Z$\sim$0.013 (standard value). 
Figure \ref{Zchanged} shows the PMS Li depletion for a
1 $M_{\odot}$ model ($Y$=0.27, $\alpha$=1.9) for the GN93 composition 
and $Z$=0.018 (model no.~16 in Table~1)
compared to our ``standard'' model (As04 mixture and $Z$=0.013); for reference,
we also plot the model with GN93 mixture and $Z$=0.013, already reported
in Fig.~\ref{mixture}. 
For the model with $Z$=0.018
the effect of an increase in the opacity -- due to the increase of $Z$ -- 
overcomes the influence of the mixture change.

\begin{figure}
\resizebox{0.5\textwidth}{!}{
 \includegraphics[width=7cm]{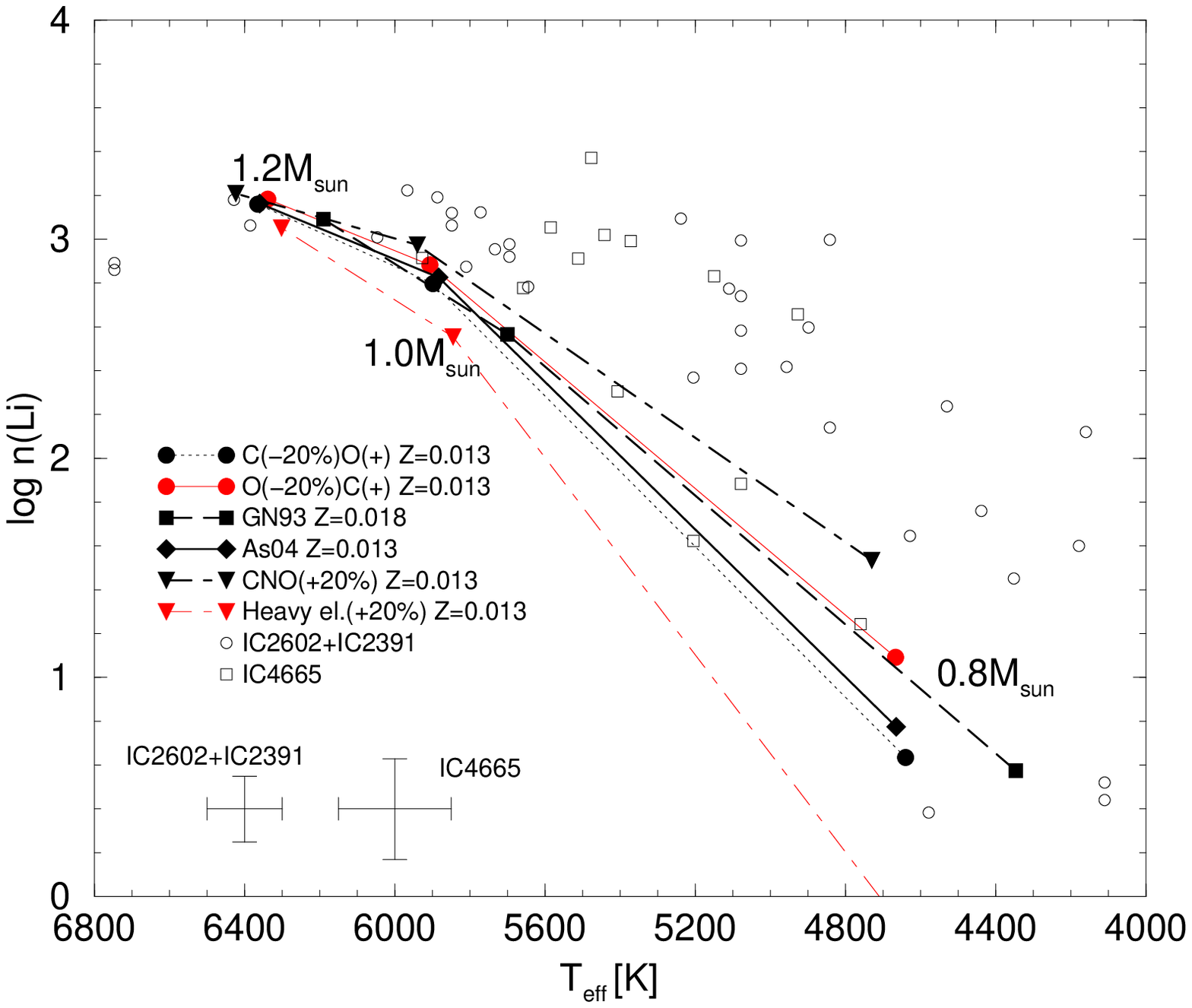}
}
 \caption{Li abundance vs.~\teff: comparison
between the observational data for young open clusters
(ages $\sim$30$\div$50 Myr) and the predictions of models. 
For the sake of clearness the calculations done for
the three selected masses for each mixture are connected by lines.
The mean observational error bars for the clusters are also shown.
 }\label{confronto}
 \end{figure}

We show in Fig.~\ref{confronto} a comparison between 
our ``standard'' model (As04 mixture),
some of the models presented in Fig.~\ref{figrisul}
(all with $Z=0.013$), the GN93 model with $Z=0.018$, 
and the observational data for three
young open clusters (ages $\sim$30$\div$50 Myr), 
in which solar-type stars are on the
ZAMS. All the clusters plotted have  [Fe/H]$\sim$0:
IC~2602 (Randich et al.~\cite{R01}), IC~2391
(Stauffer et al.~\cite{stauffer}), and IC~4665
(Mart\'{\i}n \& Montes \cite{mm97}; see also
Shen et al.~\cite{shen} for the estimate of the
metallicity).
The Li data for the clusters have been analysed 
by
Sestito \& Randich (\cite{SR05}), starting from the
equivalent widths published in the literature and using
the same method of analysis. This plot
compares the variations of our
theoretical results -- due to changes in the composition -- and Li
abundances observed in stars, 
and is not an attempt to fit the
empirical data.  For the reasons
discussed in Sect.~1, even an apparent agreement between the theoretical
predictions and the data of a cluster with a given age could be fictitious; to
claim a real agreement between theory and observations, i.e. the goodness of
the physical inputs and the macroscopic mechanisms adopted in the
calculations, the models should be simultaneously compared to data from
clusters with different chemical compositions and ages.  The plot clearly
shows how the match between theory and observations is strongly dependent on
the adopted metallicity mixture and how the situation gets worse towards lower
masses.

 %============================================== FIGURA 8
 %============================================== 
  
 %============================================== FIGURA 9
 %============================================== 
\section{Summary}

We presented an investigation of the effects of the chemical abundances on
stellar opacity and $^7$Li depletion. Within the ``standard'' 
scenario for stellar
evolution -- in which the position of 
the convective boundaries are determined
following the Schwarzschild criterion -- the theoretical predictions for the
temporal behavior of surface Li depletion are affected by various
uncertainties; among these a major role is played by radiative opacity and
therefore by the metallicity and the distribution of the single elements in
the global mixture.

Focusing on pre-main sequence evolution, where the largest amount of Li
burning is predicted by
standard theories, 
we computed stellar models for three selected masses (0.8, 1.0
and 1.2 $M_{\sun}$, with $Z=0.013$, $Y$=0.27, $\alpha$=1.9) by varying the
global mixture (Grevesse \& Noels 1993; Grevesse \& Sauval 1998; Asplund et
al. 2004; Asplund, Grevesse \& Sauval 2005), showing that the effects of the
chemical abundances must be taken into account. 
We have quantitatively computed 
the contribution of elements heavier than O for Pop.~{\sc i} models.
The comparison between models with different mixtures revealed that, at fixed
$Z$, a larger amount of Li is depleted in the presence of a larger fraction of
elements heavier than O. Therefore
these species -- although less abundant
than CNO -- play a fundamental role in stellar opacity and PMS Li
evolution.

These results have been interpreted by analysing
the contribution of the single elements to the opacity at the
temperatures and densities typical of the base of the convective envelope 
and the effects on Li depletion of single element
variations in the mixture. 

In addition, we computed models with the GN93 solar composition
in which the global
metallicity $Z$ is varied, showing that
the effect of increasing $Z$ overcomes the influence of the mixture change.
We conclude that changes in the chemical composition do affect Li
depletion. Comparisons between model predictions and observations must
take into account the element distribution in the stars. 
Moreover, the abundance determinations for open clusters should rely 
on the 3-D atmospheric models which are more consistent than the 
usually adopted 1-D models. 
None of the adopted solar mixtures
and variations in the chemical composition appear to explain the
very small amount of PMS Li depletion observed for solar-type stars.

\begin{acknowledgements}
We are extremely grateful to V. Castellani and S. Shore 
for a careful reading of
the manuscript.  We warmly thank the referee (A. Weiss) for the useful
comments. Financial support for this work was provided by the
Ministero dell'Istruzione, dell'Universit\`a e della Ricerca (MIUR)
under the scientific project ``Continuity and discontinuity in the
Galaxy formation'' (P.I.: R. Gratton).  \\
\end{acknowledgements}


\begin{thebibliography}{}
 %\bibitem[]{}
 \bibitem[1994]{AF94}Alexander, D.R.,
 \& Ferguson, J.W. 1994, ApJ, 437, 879
 \bibitem[2004]{asplund2004}Asplund, M.,
 Grevesse, N., Sauval, A.J., Allende Prieto, C., \& Kiselman, D. 2004, A\&A, 417, 751
 \bibitem[2005]{asplund2005}Asplund, M.,
 Grevesse, N., \& Sauval, A.J. 2005, in ``Cosmic Abundances as
 Records of Stellar Evolution and Nucleosynthesis", eds. F.N. Bash,
 \& T.J. Barnes, ASP Conf. Series, 336, 25
\bibitem[2005] {Bahcall} Bahcall, J.N., \& Serenelli, A.M. 2005, ApJ, 626, 530
 \bibitem[2003]{bensby}Bensby, T., Feltzing, S., \& Lundstr\"om, I.
2003, A\&A, 410, 527
 \bibitem[2004]{cariulo} Cariulo, P., Degl'Innocenti, S., \& Castellani,
   V. 2004, A\&A, 424, 927
\bibitem[2004]{carretta04}Carretta, E., Bragaglia, A., Gratton, R., \& Tosi,
M. 2004, A\&A, 422, 951
 \bibitem[1999]{caste99}Castellani, V., Degl'Innocenti, S., \& Marconi, M.
 1999, MNRAS, 303, 265
\bibitem[2000]{chen}Chen, Y.Q., Nissen, P.E., Zhao, G., Zhang, H.W.,
\& Benoni, T. 2000, A\&AS, 141, 491
 \bibitem[1989]{chieffi}Chieffi, A., \& Straniero, O. 1989, ApJS, 71, 47
 \bibitem[1997]{ciacio97}Ciacio, F., Degl'Innocenti, S.,
 \& Ricci, B. 1997, A\&AS, 123, 449
 \bibitem[1997]{dm97}D'~Antona, F., \& Mazzitelli, I. 1997, MmSAIt, 68, 807
\bibitem[2002]{friel02}Friel, E., Janes, K.A., Tavarez, M., et al.~2002,
AJ, 124, 2693
\bibitem[1991]{grevesse91a}Grevesse, N. 1991, A\&A 242, 488
\bibitem[1991]{grevesse91}Grevesse, N., \& Anders, E. 1991,
 in ``Solar Interior and Atmosphere'', Tucson, AZ, University of Arizona
 Press, p.1227
 \bibitem[1993]{grevesse93}Grevesse, N., \&
 Noels, A. 1993 in ``Origin and Evolution of the Elements'', eds.
 N. Prantzos, E. Vangioni--Flam, \& M. Cass\'e, Cambrdige
 University Press, Cambridge, p.~15
 \bibitem[1998]{grevesse98}Grevesse, N., \& Sauval, A.J. 1998, SSRv, 85, 161
 \bibitem[1996]{OPAL}Iglesias, C., \& Rogers, F.J.
 1996, ApJ, 464, 943
\bibitem[2000]{jef00}Jeffries, R.D. 2000 in ``Stellar Clusters and Associations:
Convection, Rotation, and Dynamos'', eds. R. Pallavicini, G.
Micela, \& S. Sciortino, ASP Conf. Ser. 198, p.~245
 \bibitem[2006]{jef_cast04}Jeffries, R.D. 2006, in ``Chemical Abundances and Mixing in Stars in the Milky Way
  and its Satellites'', eds. L. Pasquini,
 \& S. Randich, ESO Astrophysic Symposia, in press
 \bibitem[2003]{luridiana}Luridiana, V., Peimbert, A., Peimbert, M.,
 \& Cervi\~no, M. 2003, ApJ, 592, 846
\bibitem[1997]{mm97}Mart\'{\i}n, E., \& Montes, D. 1997, A\&A, 318, 805
 \bibitem[1988]{nissen}Nissen, P. 1988, A\&A, 199, 146
\bibitem[2004]{nord}Nordstr\"om, B., Mayor, M., Andersen, J., et al. 
2004, A\&A, 418, 989
 \bibitem[2004]{olive}Olive, K.A., \& Skillman, E.D. 2004, ApJ, 617, 29
 \bibitem[1998]{pagel}Pagel, B.E.J., \& Portinari, L. 1998, MNRAS, 298, 747
 \bibitem[2000]{Palla}Palla, F. 2002
  in ``Star Formation and the Physics of Young Stars", eds.
 Bouvier, \& J.-P. Zahn, EDP Sciences, Aussois
  (France) EAS Publications Series, Vol. 3 p. 111
\bibitem[2005]{piau05}Piau, L. 2005, submitted to ApJ, astro-ph/0511402
\bibitem[2002]{ptc02}Piau, L., \& Turck-Chi\`eze, S. 2002, ApJ, 566, 419
 \bibitem[2003]{piau}Piau, L., Randich, S., \& Palla, F. 2003, A\&A, 408, 1037
 \bibitem[1997]{pinso}Pinsonneault, M. 1997, ARAA, 35, 557
 \bibitem[1999]{Potekin99}Potekhin, A. Y. 1999, A\&A, 351, 787
\bibitem[2006]{R05_cast04}Randich, S. 2006, in
``Chemical Abundances and Mixing in Stars in the Milky Way
 and its Satellites'', eds. L. Pasquini, \& S. Randich, ESO Astrophysic
Symposia, in press
\bibitem[2001]{R01}Randich, S., Pallavicini, R., Meola, G.,
Stauffer, J.R., \& Balachandran, S.C. 2001, A\&A, 372, 862
\bibitem[2006]{R06}Randich, S., Sestito, P., Primas, F.,
Pasquini, L., \& Pallavicini, R. 2006, in press, astro-ph/0601239
\bibitem[2001]{RO01} Rogers, F.J., 2001,  Contributions to plasma physics,
  V41(N2-3), 179 
 \bibitem[2006]{schuler05}Schuler, S.C., Hatzes, A..P, King, J.R.,
Kuerster, M., \& The, L.-S. 2006, AJ, 131, 1057
\bibitem[2005]{SR05}Sestito, P., \& Randich, S. 2005, A\&A, 442, 615
\bibitem[2003]{sestito03}Sestito, P., Randich, S., Mermilliod, J.-C.,
\& Pallavicini, R. 2003, A\&A, 407, 289
\bibitem[2005]{shen}Shen Z.-X., Jones, B., Lin, D.N.C., Liu, X.-W.,
\& Li, S.-L. 2005, ApJ, 635, 608
 \bibitem[2000]{siess}Siess, L., Dufour, M., \&
 Forestini, M. 2000, A\&A, 358, 593
 \bibitem[2004]{stahler}Stahler, S.W., \& Palla, F. 2004, ``The Formation
 of Stars'', New York, Wiley
\bibitem[1989]{stauffer}Stauffer, J., Hartmann, L.W., Jones, B.F., \& McNamara, B.R. 1989, ApJ, 342, 285
 \bibitem[1994a]{swenson94a}Swenson, F.J, Faulkner, J, Iglesias,
   C.A., Rogers, F.J., \& Alexander, D.R. 1994a, ApJ, 422, L79
 \bibitem[1994b]{swenson94b}Swenson, F.J, Faulkner, J, Rogers, F.J., 
 \& Iglesias, C.A. 1994b, ApJ, 425, 286
 \bibitem[1994]{thoul}Thoul, A.A., Bahcall, J.N., \&
 Loeb, A. 1994, ApJ, 421, 828
 \bibitem[1998]{Turcotte} Turcotte, S., \& Christensen-Dalsgaard, J.
 1998, Space Sci. Rev., 85, 133
\bibitem[1998a]{turcotte98a} Turcotte, S., Richer, J., \& Michaud, G.
1998a, ApJ, 504, 559
\bibitem[1998b]{turcotte98b} Turcotte, S., Richer, J., Michaud, G.,
Iglesias, C., \& Rogers, F. 1998b, ApJ, 504, 539
 \end{thebibliography}
 \end{document}